\documentclass[useAMS,usenatbib]{mn2e}
\usepackage{graphicx}
\usepackage{multirow}

\def\etal{\emph{et\,al.}}

\title[Fibre MOS sky subtraction]{Sky subtraction at the Poisson limit
with fibre-optic multi-object spectroscopy}

\author[Sharp \& Parkinson]{R. Sharp$^{1}$\thanks{E-mail: rgs@aao.gov.au} and Parkinson H.$^{2}$\\
$^{1}$Anglo-Australian Observatory, P.O. Box 296, Epping, NSW, 1710, Australia\\
$^{2}$Institute for Astronomy, Edinburgh University, Royal Observatory,
Blackford Hill, Edinburgh, EH9 3HJ, UK}

\begin{document}

\date{Accepted YYYY MM NNM Received YYYY MMM NN; in original form YYYY MMM NN}

\pagerange{\pageref{firstpage}--\pageref{lastpage}} \pubyear{2002}

\maketitle

\label{firstpage}

\begin{abstract}
We report on the limitations of sky subtraction accuracy for long
duration fibre-optic multi-object spectroscopy of faint astronomical
sources during long duration exposures.  We show that while standard
sky subtraction techniques yield accuracies consistent with the
Poisson noise limit for exposures of 1\,hour duration, there are large
scale systematic defects that inhibit the sensitivity gains expected
on the summation of longer duration exposures.  For the AAOmega system
at the Anglo-Australian Telescope we identify a limiting systematic
sky subtraction accuracy which is reached after integration times of
4-10\,hours.  We show that these systematic defects can be avoided
through the use of the fibre nod-and-shuffle observing mode, but with
potential cost in observing efficiency.  Finally we demonstrate that
these disadvantages can be overcome through the application of a
Principle Components Analysis sky subtraction routine.  Such an
approach minimise systematic residuals across long duration exposures
allowing deep integrations.

We apply the PCA approach to over 200\,hours of on-sky observations
and conclude that for the AAOmega system the residual error in long
duration observations falls at a rate proportional to $\tau^{-0.32}$
in contrast to the $\tau^{-0.5}$ rate expected from theoretical
considerations.  With this modest rate of decline, the PCA approach
represents a more efficient mode of observation than the
nod-and-shuffle technique for observations in the sky limited regime
with durations of 10-100\,hours (even before accounting for the
additional signal-to-noise and targeting efficiency losses often
associated with the N+S technique).

This conclusion has important implications for the observing
strategies of the next generation of fibre-optics redshift surveys
with existing facilities as well as design implications for
fibre-optic systems destined for new facilities.  It argues against
the use of the inherently inefficient nod-and-shuffle technique for
faint object fibre-optic survey spectroscopy.
\end{abstract}


\section{Introduction}
High multiplex multi-object spectrographs have been key to many core
astronomical investigations since their routine operation began over
three decades ago.  However, even with the great advances made in our
understanding of the Universe via large scale spectroscopic surveys
(such as those previously under taken at the AAT with the 2dF instrument:
2dFGRS - \citet{2dFGRS}, 2QZ - \citet{2QZ}, WiggleZ - \citet{WiggleZ}, GAMA -
\citet{GAMA}), or closer to home with stellar surveys (such as the
UKST RAVE project - \citet{RAVE}), there remains an ever increasing
desire for more powerful survey machines.  As astronomy enters the era
of the Extremely Large Telescopes (ELT) it is appropriate to pause and
assess the techniques and methodologies available to modern
instrumentation in order to fully utilise the valuable photons
gathered by these gargantuan machines.

In the same vein, older facilities strive to identify new modes of
operation. These stalwarts of the past two decades of research must
find the best modes of operation to support the ambitions of the new
generation of facilities, whose availability will be restricted at
best.  It is critical that we take time to review the options and
identify the roles each instrument must play to maximise the overall
return on our significant investments, both material and intellectual.

There are many subtleties to the process of spectrograph design and
that discipline is not considered here (for an introduction see
\citet{Hearnshaw09}) but the problem of MOS system design is in
essence that of identifying the most expedient reformatting of the
telescope focal plan in order to deliver dispersed light to the
detector system.  Historically three methods have been employed,
objective prism surveys, focal plane slit-mask slit-spectroscopy and
fibre-optic systems.  In any high multiplex spectroscopic system the
correct mode of observations is governed by a number of competing
survey demands:\\

\vspace{-0.25cm}
\noindent{\it Target Multiplex -} Maximising the total number of
science targets which can be simultaneously observed.\\

\vspace{-0.25cm}
\noindent{\it Signal-to-noise -} Minimising the required integration
time to reach a predefined signal-to-noise level within the elements
of each spectrum.\\

\vspace{-0.25cm}
\noindent{\it Spectral quality -} Minimising the systematic errors
within each spectrum.\\

\vspace{-0.25cm} Systematic defects can of course be considered as an
additional noise source.  However while noise associated with a
Poisson process, such as the rate of photon arrival at the detector,
can be suppressed as the square root of the observation time,
systematic defects are typically removed at a slower rate.

In this paper we focus on the use of fibre-optic systems and
specifically on an optimal strategy for the accurate subtraction of
the OH air-glow lines which plague optical spectroscopy in the
wavelength region above $\sim$6000\AA.  Inaccurate subtraction of the
OH lines is the dominant source of systematic error in sky limited
fibre spectroscopy at the red end of the optical spectrum.  There are
many discussions of the OH problem in the literature and we refer the
reader to the recent review of \citet{Ellis08}.  In this work we
consider the merits of two novel techniques alongside the traditional
dedicated sky fibre method.  We investigate the use of the {\it
nod-and-shuffle} mode of observation \citep{Glazebrook01} coupled with
the use of a high multiplex fibre-optic multi-object spectroscopy
system. We also investigate the application of a Principle Components
Analysis (PCA) to improve sky subtraction, following the techniques of
\citet{Kurtz00} and \citet{Wild05} as implemented by \citet{Parkinson}
and outlined in \S\ref{pca overview}.  We use data sets from the
AAOmega facility at the 3.9m Anglo-Australian Telescope (AAT) to
demonstrate the results.

\section{Observational data sets}
\label{observationaldata}
The example spectra throughout this work where taken with the the
AAOmega spectrograph (\citet{Saunders04}, \citet{Sharp06}) at the
Anglo-Australia Telescope (AAT). The fibre feed from the 2dF robotic
positioner to the AAOmega spectrograph uses 140$\pm$2\,$\mu$m core
optical fibre
\footnote{Polymicro FBP140168196 - Broadband Polyimide coated
140$\pm$2\,$\mu$m core, 168$\pm$2\,$\mu$m, cladding 196$\pm$3\,$\mu$m
buffer.} which project to $\sim$2\,arcsec at the focal plane of the
2dF prime focus corrector.

Observations are shown from the red arm of the dual beam AAOmega
system, using the 385R Volume Phase Holographic (VPH) grating.  Three
data sets from two programs are used for the purposes of
illustration. Details are given in Table~\ref{data sets}.  In the
first instance moderate signal spectra from the GAMA project
\citep{GAMA} are used.  Secondly, in order to track long term
systematic error trends $\sim$200\,hours of low signal, essentially
sky limited, observations from the WiggleZ dark energy project
\citep{WiggleZ} are used.  Finally to demonstrate the nod-and-shuffle
technique, a series of deep N+S frames, taken as part of the
completeness estimation process for the WiggleZ project, are used.
Data were taken using both the 5700\AA\ and 6700\AA\ dichroic beam
splitters.
Example spectra are shown in units of CCD counts converted to
electrons via the measured gain parameter.

\subsection{Data processing}
\label{dataprocessing}
The data used in the following analysis has been processed using the
\texttt{2dfdr} data analysis package.  All data was processed using
the standard settings for low resolution extra-galactic data.  Sky
subtraction has been performed using three techniques.\\

\vspace{-0.25cm}
\noindent{\it Dedicated sky fibres -} Twenty-five fibres where
allocated to known blank sky positions.  The spectra from these fibres
are median combined to create a master sky spectrum.  This master sky
spectrum is then scaled for subtraction from each science spectrum
using the OH lines.\\

\vspace{-0.25cm}
\noindent{\it Nod-and-shuffle (N+S) -} Observations where performed in
this mode \citep{Glazebrook01} using a pair of fibres allocated to
each science target (Cross Beam Switching, CBS).  These observation
are processed in a manner similar to standard AAOmega data with the
exception that sky subtraction is achieved by direct subtraction of
the raw CCD pixel intensities for the A and B telescope nod positions
prior to spectral extraction.\\

\vspace{-0.25cm}
\noindent{\it Principle components analysis (PCA) -} The final method
used has been to perform the standard {\it dedicated sky fibres}
reduction but once complete the data was processed using the a
Principle Components Analysis sky subtraction routine as implemented
for the GAMA survey by \citet{Parkinson}.\\

\vspace{-0.25cm} The sky spectrum recorded during each integration is
used as the normalisation when considering the residual sky
subtraction error in any given data set.

\subsection{Overview of nod-and-shuffle}
The {\it nod-and-shuffle} technique has been shown to provide Poisson
limited performance in slit spectroscopy \citep{Glazebrook01} with
perhaps the best known example of the results achievable being the
deep spectroscopy of high redshift early-type galaxies from the Gemini
Deep Deep Survey \citep{gdds}.  A nod-and-shuffle observing mode for
fibre-optic MOS had been provided with the 2dF system at the AAT but
the significant practical inefficiencies of the implementation
hampered observations.  However, the AAOmega spectrograph was
commissioned with 200 and 400 fibre nod-and-shuffle modes readily
available.

For nod-and-shuffle observations, the telescope is nodded
back-and-forth between the target and adjacent blank sky at a
frequency of 0.5-2\,Hz. During the the telescope {\it nod} the charge
accumulated on the CCD is transferred without loss ({\it shuffled}) to
a storage area and a second region of the CCD is exposed to the sky
flux now illuminating the fibre.  This process is repeated a number of
times before the CCD is readout.  This mode allows quasi-simultaneous
observations of source and sky, through an identical light path.
Allocating a pair of fibre to each source and nodding the source
between each fibre in the pair mitigates the implicit 50\% on-source
duty cycle at the expense of target multiplex.

As will be seen below fibre-optic nod-and-shuffle performs well.
However, as well as significant practical inefficiencies such as
telescope slew-and-settle time and reduction in target multiplex the
nod-and-shuffle technique also introduces a significant reduction in
the signal-to-noise ratio achieved, with respect to the dedicated sky
fibre approach, for a finite observing time (see \S\ref{Appendix}).

\subsection{Overview of Parkinson \etal\ PCA implementation}
\label{pca overview}
The use of a PCA approach for sky subtraction was first demonstrated
by \citet{Kurtz00}, although for that first implementation classical
long-slit spectroscopic data was used to simulate the performance of
the procedure.  The first true test of the PCA approach with fibre
spectroscopy was reported by \citet{Wild05} using data from SDSS.
However, \citet{Wild05} focused on {\it cleaning up} relatively
shallow spectroscopy.  A goal of many fibre MOS projects has been to
achieve more sensitive observations.  Indeed, the WiggleZ Dark Energy
survey \citep{WiggleZ} utilized a modified version of the
\citep{Wild05} PCA code to aid in the identification of faint
(r(AB)$<$22.5) $z>0.5$ emission-line galaxies whos characteristic
emission lines are otherwise lost in the dense OH-forest at
wavelengths beyond $\sim$7000\AA.

While early implementations have been successful, the PCA sky
subtraction technique has seen little use over the past decade.  There
are numerous underlying reasons for this lack of interest in a
potentially powerful technique:\\

\vspace{-0.25cm}
- Much early work was undertaken at relatively blue
wavelengths where the OH sky-line forest is spares and lines are of
only moderate intensity.  With the advent of modern large format CCD
system which retain high quantum efficiency out to 1\,$\mu$m and
beyond an important new spectroscopic window is opening up to fibre
spectroscopy.\\

\vspace{-0.25cm}
- Slit systems report excellent results from accurate
slit mapping in 2D, so the technique has not been required for
long-slit or slit-mask spectroscopy systems\footnote{However,
systematics do remain, even after correction with a suitable slit
function (commonly due to flexure in the spectrograph).  A detailed
measurement with long duration observations has not been performed.}.
While this will remain true in the coming decade, there is no shortage
of current or planned fibre MOS systems, largely due to the relative
ease of implementation for fibre MOS systems accessing large fields of
view.\\

\vspace{-0.25cm} - Relatively bright targets have traditionally been
observed with fibre based MOS systems.  This is in part due to the
historically somewhat marginal results that have been achieved on
fainter sources due to poor control of systematic spectral defects in
long duration exposures.  It is these defects that the PCA process
addresses.\\

\vspace{-0.25cm} - Most fibre MOS systems report residual error
estimates close to the Poisson residuals expected for routine
observations and so there has been little need for improvement.  We
demonstrate that the application of the PCA approach will allow
significantly more difficult observations to become routine.

The approach to PCA sky subtraction implemented for the AAOmega system
at the AAT varies from the previous works and is described in detail
in \citet{Parkinson}.  The implementation used in this work was a
standalone processing script which ran on the reduced data products
from the \texttt{2dfdr} data reduction pipeline.  The process has now
been incorporated into a pipeline processing option within the body of
the reduction code.  A brief overview is given here:\\

\vspace{-0.25cm} - Each target field is observed with $\sim$400
science fibres. Of these, 20-25 are allocated to blank sky positions
across the field to provide night sky spectra free from astronomical
sources.\\

\vspace{-0.25cm} - The data are processed in the usual manner using
the \texttt{2dfdr} data reduction software.  To perform a first order
sky subtraction a master sky spectrum is created from a median
combination of the dedicated blank sky spectra.  This spectrum is then
scaled and subtracted from each science frame using the median flux in
strong OH night sky lines to determine the required scaling.\\

\vspace{-0.25cm} - Each valid spectrum (i.e.\ source and sky fibres
are used, but bright calibration sources and unused fibres are
discarded) is then continuum subtracted using a rolling median filter
to remove large scale source structure.\\

\vspace{-0.25cm} - A set of eigen vector spectra is calculated using a
Principle Components Analysis for the $\sim$400 sky subtracted
spectra.\\

\vspace{-0.25cm} - Each science spectrum is then sky subtracted using
a linear sum of the thirty most significant principle components.
\newline

The main departures of the GAMA project PCA implementation of
\citet{Parkinson} from that of \citet{Wild05} for data from SDSS
are:\\

\vspace{-0.25cm} - The eigen vectors and values are generated on a
frame-by-frame basis, rather than from a large data base of spectra.
This addresses many of the major sources of poor sky subtraction
residuals (\ref{Appendix-problems}).\\

\vspace{-0.25cm} - Object and sky spectra are used to generate the
components.  This clearly breaks down if numerous objects have very
strong structure at common wavelengths (e.g.\ M-star contamination of
a galaxy survey) or if many similar objects are found within a given
frame (i.e.\ during galactic survey work).\\

\vspace{-0.25cm} - Spectral features are not masked since they are not
know {\it a priori}.  Indeed, enhancing the detectability of such
features within a strong foreground signal is the goal of this work.\\

\vspace{-0.25cm} - With only $\sim$400 input spectra for each frame,
only a small number of eigenspectra can be use to perform the sky
subtraction.  In the work present here thirty components have been
used.  This number was selected for initial investigations as reported
here by inspection of the resulting spectra.  A full investigation of
the methodologies for parametrisation the eigenvector space from which
components are drawn is underway.

\subsubsection{Continuum concerns}
Sky and source continuum presents a problem.  In principle the
residual sky continuum should represent a separate set of distinct
components in the eigenvector set produced by the PCA algorithm.  In
practise however, the residual OH and continuum signals and the source
spectrum do not represent orthogonal basis sets and so intermixing
between source spectral components in such data is inevitable.  It
will then not be possible to adequately represent any individual sky
subtraction residual pattern using a reconstruction from a limited set
of these eigenvectors.

\citet{Kurtz00} deal with the problem via a two stage fitting process
which has the unfortunate side effect of invalidating the spectrum for
measurements of the continuum properties.  \cite{Wild05} rely on the
fact that each SDSS spectrum has had a preliminary sky subtraction to
remove sky continuum components prior to their correction for OH line
residual via the PCA.  Object continuum is removed via median
filtering each spectrum in wavelength.  In this regard our approach is
identical to that of \citet{Wild05}.

Ultimately it may be possible to assess the selection of eigenvectors
and weighting based on those that preferentially contribute to
residual correction in the blank-sky fibres spectra (perhaps
requiring an increased number of dedicated sky spectra per
observation).  We have yet to investigated this aspect of problem.  In
short, we expect our approach to provide a reliable result for low
intensity source spectra taken on moonless nights (so that the first
order sky correction is adequate).  The primary interest should be
focused on the red end of the spectrum where the error budget is
dominated by OH subtraction errors and the source distribution should
be such that individual spectra have little in common in the observed
frame.  A small number of high signal, or repetitive spectra (such as
calibration stars) should be clipped from the generation of the
initial eigenvector set.

For work in which absolute spectral fidelity at the highest level is
of paramount importance at the expense of significant survey
efficiently - not typically the primary motivation in work such as a
wide area galaxy redshift survey - one may require the {\it
nod-and-shuffle} technique.

\subsection{Sky subtraction accuracy}
\label{skysubdefinition}
Example sky spectra as observed by the red arm of the AAOmega
spectrograph are given in Fig.~\ref{skyspec}.  Two horizontal levels
are marked.  These indicate the residual Poisson error, expressed as a
fraction of the local sky spectrum, which would be expected under the
assumption of pure Poisson noise in the spectrum of a 1\,hour sky
observation.  The majority of the spectral region considered here is
limited to a residual noise level between 1-3\% of the sky signal in
each spectral pixel.

We require a metric with which to determine the sky subtraction
accuracy achieved in any given observational data set.  The figure of
merit we use is the {\it local error} which we estimate as follows.

A smoothed spectrum is constructed using a sliding square filter (with
smoothing length of $\pm$100\,pixels) and computing a 5-$\sigma$
clipped mean at each pixel.  For each pixel the rms intensity within
the smoothing window is taken as an estimate of the {\it local sky
subtraction accuracy}.

To confirm the validity of this error estimator a number of
simulations are undertaken. Random realisation of noise spectra are
generated with pixel intensities drawn from two distributions.  The
first is a Normal distribution with width set to that indicated by the
variance array information which is generated for each spectral pixel
during data reduction.  A second realisation is generated from a
normal distribution with width set by the {\it local error estimate}
defined above.  A thousand random realisation of each distributions
are generated for each spectral pixel and binned histograms of these
model pixel intensities are compared to the pixel intensities for the
observational spectra after subtraction of the smoothed spectrum.
Inspection of the these residual error histograms indicates\\

\noindent- The variance estimates from the data reduction process are
an accurate description of the true noise properties of the data, with
perhaps some small evidence for departure (likely due to small errors
in the gain and CCD read-noise, N$_{rd}$, estimation).\\

\noindent- The {\it local error estimator} reliably reproduces the
distribution of observed pixel indensities and hence is a valid
representation of the residual scatter in the data.

\section{Example spectra}
Sky subtraction accuracy is estimated for the experimental data sets
detailed in \S\ref{observationaldata} using the three subtraction
methods presented in \S\ref{skysubdefinition}.  The results are given
in Table~\ref{skysubresults}.  The individual data sets are discussed
below.

\subsection{GAMA survey - moderate source intensity data}
\begin{figure}
\includegraphics[width=87mm]{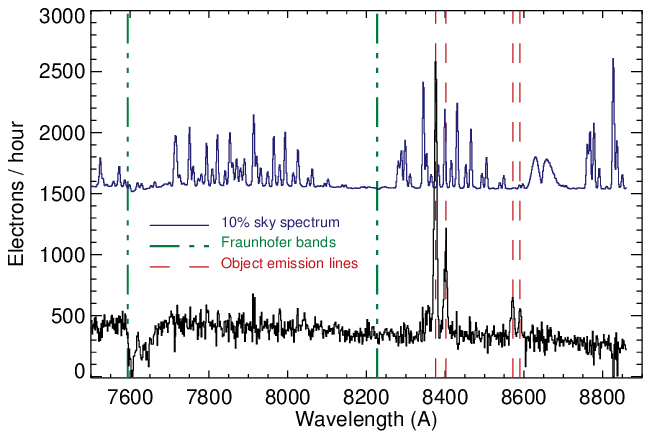}\\
\includegraphics[width=87mm]{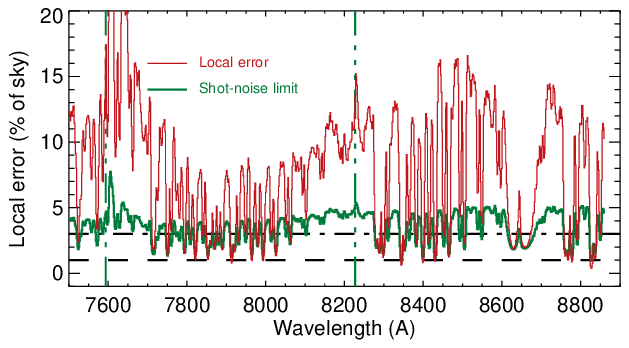}
\caption{\label{GAMA obj} An object spectrum from the GAMA survey is
shown to illustrate the sky subtraction accuracy for moderatly bright
sources using the dedicateted sky subtrcation technique.
\newline
The upper spectrum is composed of 3$\times$1200\,sec exposures of
1\,hour on source using 24 dedicated sky fibres to construct the
master sky spectrum used for sky subtraction.  The emission line
source shown (H$\alpha$/[N\textsc{ii}]/[S\textsc{ii}]) is at a
redshift of z=0.276.  The source catalogue magnitude is r(AB)=19.17.
The plotting range is chosen to show regions of strong atmospheric OH
and O$_{\rm2}$ emission, sky line free continuum regions and two
Fraunhofer telluric absorption bands.
\newline
The lower plot shows the local error in the spectrum, estimated as
discribed in \S~\ref{skysubdefinition}.  The error is presented as a
percentage of the local sky spectrum at each wavelength.  The Poisson
limit determined from pixel varaince array during data reduciton is
indicted by the thick line.  The Poisson limit is approached in the
vicinity of the strong emission lines.  Between the emission features,
the residual is dominated by Poisson noise from the source spectrum.}
\end{figure}

Observations from the GAMA project \citep{GAMA,Baldry09} are shown in
Fig.~\ref{GAMA obj}.  The source presented has reported SDSS
magnitudes of $r$(AB)=19.17 and is found to be at redshifts of
$z$=0.276.  The spectrum is composed of 3$\times$1200\,sec
integrations, and was processed with the use of 24 dedicated sky
fibres to provide the master sky spectrum for subtraction.  This
master sky spectrum is shown scaled to 10\% for comparision.  While
the sky subtraction does not hamper the identification of redshift for
the source, there are systematic residuals in the data.

The lower panel in Fig.~\ref{GAMA obj} shows the {\it local error
estimate} as defined in \S~\ref{skysubdefinition}.  The local error
varies between close to the Poisson noise limit and over 10\% of the
local sky spectrum with an rms 4.67$\pm$0.6 over the full spectrum.

In the absence of a Poisson component from the target source
continuum, the local error estimate is consistent with the variance
estimation derived from the pixel intensities during data reduction.

\begin{figure}
\includegraphics[width=80mm]{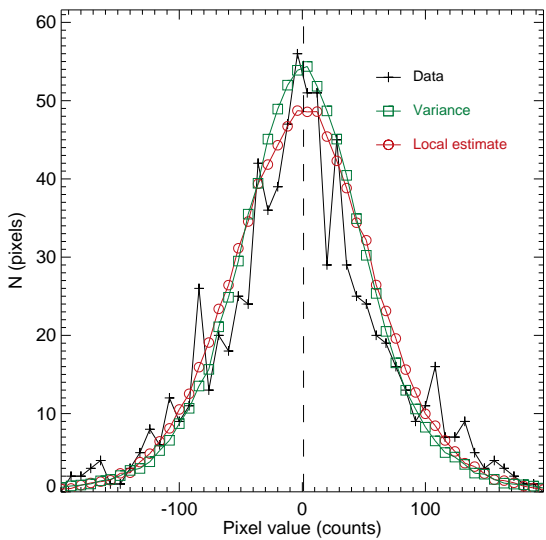}
\caption{\label{GAMA res1} The residual error per pixel is assessed
for the spectrum presented in Fig.~\ref{GAMA obj}.  The local error
estimate reproduces the distribution of data values well.}
\end{figure}

The residual error histogram for the spectrum is shown in
Fig.~\ref{GAMA res1}, confirming that the local error estimate is
consistent with the residual error in the source spectra which is in
turn broadly consistent with the spectral error expected from the data
reduction pixel variance.

\subsection{WiggleZ data - sky limited data}
\begin{figure}
\includegraphics[width=85mm]{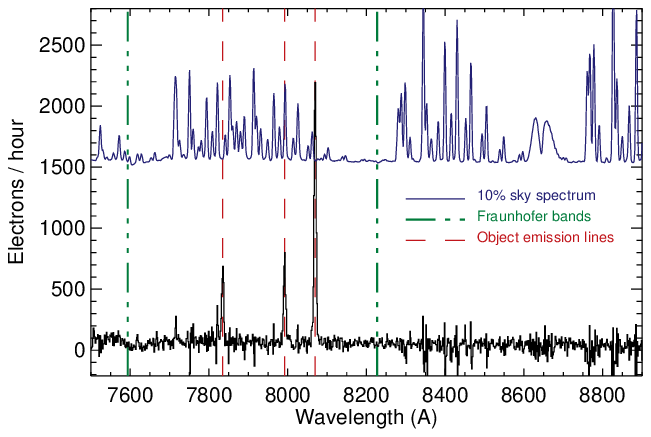}\\
\includegraphics[width=85mm]{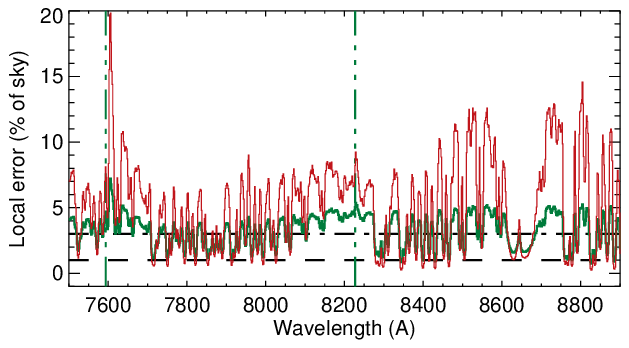}
\caption{\label{WiggleZ obj} Fig.~\ref{GAMA obj} is repeated for a
source from the WiggleZ survey.  WiggleZ targets are selected to be
$r$(AB)$>$21 emission line galaxies and so each spectrum is free from
contamination from source continuum in most pixels.  The source shown
has $r$(AB)=21.63 and an emission line redshifts of z=0.6117. The
spectrum is a 3$\times$1200\,sec observation.  The lack of significant
objects continuum accounts for the improved local error estimates for
the WiggleZ spectra over that shown in Fig.~\ref{GAMA obj}.  The local
error in the spectrum approached the Poisson limit in the vicinity of
strong OH emission features.}
\end{figure}

Data from the WiggleZ survey \citep{WiggleZ} is in a different regime
from that of the GAMA survey.  Since WiggleZ sources are 20 $<$
$r$(AB) $<$ 22.5 high equivalent width [O\,\textsc{ii}] emission line
sources there is little source continuum to introduce increased noise
to the error statistic.  Fig.~\ref{WiggleZ obj} shows such a WiggleZ
survey source.  The local error is close to the sky limited case
(i.e.\ the limit expected from the pixel intensity variance
estimates).

Since WiggleZ spectra are largely free of continuum a stacking
analysis of the all 352 science fibre spectra in this frame should
produce an improvement in the local error estimate which scales as
$\sqrt{\rm{N}}$ for the number of fibres stacked.  A clear departure
from this scaling is seen in the upper panel of Fig.~\ref{WiggleZ
sky}.  This failure to follow the $\sqrt{\rm{N}}$ suppression is the
result of systematic errors in the sky subtraction.

On applying the PCA correction to the sky subtracted spectra and
repeating the stacking analysis the local error estimate is seen to
improve markedly.  This control of the build-up of systematic error is
easily seen in the residual error histogram of Fig.~\ref{WiggleZ
hist}.  Without the PCA correction the local error estimator is
clearly very different from the expected error distribution (based on
pixel variance estimates from data reduction).  With the application
of the PCA correction prior to stacking, the local error estimator and
variance data are largely in agreement.

\subsection{WiggleZ Deep - Nod-and-Shuffle data}
To compare the previous results to observations taken with the
nod-and-shuffle observing technique we use deep observations taken,
for quality control purposes, as part of the WiggleZ project.  A
6\,hour on source observation is available comprising three
observations (taken over three nights) of 3$\times$2400\,sec.  The
data was observed in the Cross-Beam-Switching (CBS) mode with a pair
of fibres allocated to each target.  The observations where taken in
the conventional AAOmega N+S mode with half of the available science
fibres masked in order to provide the requisite CCD storage space
(rather than the recently implemented mini-shuffling mode discussed
later and in \citet{minishuffle}) .

\begin{figure}
\includegraphics[width=85mm]{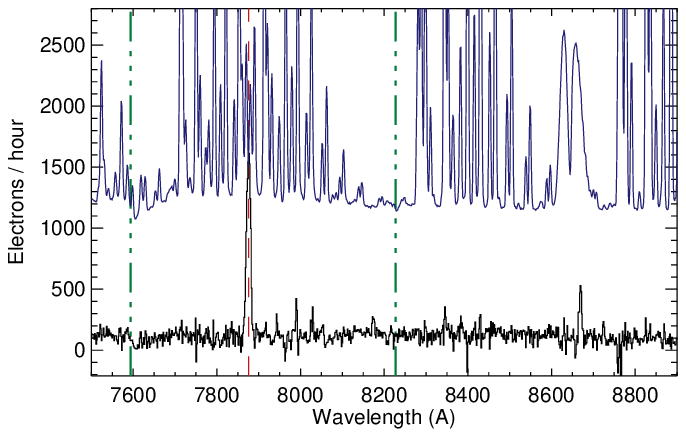}\\
\includegraphics[width=85mm]{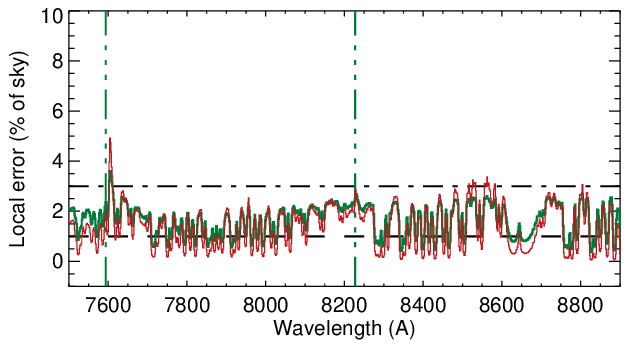}
\caption{\label{wigglez ns3} The full 6\,hour WiggleZ N+S observation
of an emission line source is shown.  The source has r(AB)=22.26 and
the redshift indicated by the [O\,\textsc{ii}] emission line is
z=0.851.  Using a CBS fibre pair the observation comprises 6\,hours on
sky and 6\,hours on source.}
\end{figure}

A representative spectrum for a typical source ($r$(AB)=22.26,
$z$=0.851) is presented in Fig.~\ref{wigglez ns3}.  The local error
estimator indicates that the observations are indeed close to the
Poisson noise limit expected for the N+S observing process.

\section{Stability with long duration exposures} 
Observations of faint sources require long duration exposures to build
up the required signal-to-noise ratio within each spectrum.  In the
limit of pure Poisson-noise in the photon arrival rate,
signal-to-noise is expected to build as the square-root of the
observation time.  However, systematic defects in the spectrum, most
commonly from systematic failure of the sky subtraction, will reduce
the efficiency of this accrual of signal-to-noise.  The common causes
of these systematic defects in fibre spectroscopy are considered in
appendix~\ref{Appendix-problems}.  Determining the point at which each
sky subtraction methodology becomes limited by systematic defects is
key to selecting the most efficient observational strategy to achieve
ones goals.

In the absence of an ideal data set we mimic long duration exposures
using two methods.  First we consider the effects of stacking fibre
spectra from within a single MOS frame.  Secondly we investigate
stacking individual fibres from across multiple independent
observations.  The former is not ideal as there may be strong
correlated noise patterns between fibres in a single frame which would
average out over many frames.  The later is difficult in practise as a
suitable data set, which places a single fibre on blank sky across
many independent observing frames, is not usually available.

The residual error as a function of time for a single frame stacking
process are shown in Fig.~\ref{noPCA stack}. Single frames from the
GAMA, WiggleZ and WiggleZ N+S data sets presented previously are used.
Additionally, in order to construct a single fibre multi-frame stack
we use the properties of the WiggleZ survey data to construct a stack
of $+$200 individual fibre spectra from data taken throughout 2009
(Fig.~\ref{PCA stack}).  In each case stacked spectra are created with
a range of values of N, the number of fibres used.  For each value of
N, 100 random realisations of the stacked spectrum are created and the
local error and its rms scatter are plotted against the number of
spectra used in the stack.  Each stack of N fibres simulates an
exposure of N\,hours on blank sky.  The rate of reduction in residual
error with time is then compared to the $\sqrt{\rm{N}}$ rate of
decline expected from pure Poisson statistics.

It is clear from Fig.~\ref{noPCA stack} that for both the data sets
which use purely the {\it dedicated sky fibre} subtraction, a
systematic noise floor is reach after $\sim$10\,hours.  With the
application of the PCA procedure to the data prior to stacking there
is no indication of this systematic noise floor until one reaches
$\sim$100\,hours of simulated observation.

The N+S observations show little evidence for a noise floor at the
limits of the available exposure time, however the residual error
appears to decline at a rate slower than the $\sqrt{\rm{N}}$ Poisson
limit, rather the rate of decline estimated from the data set is
N$^{-0.38}$.

To investigate the decline rate further we perform a single fibre
stacking across all 219 observations available from the 2009 WiggleZ
project observing program at the AAT.  The results of this long
duration exposure stacking are shown in Fig.~\ref{PCA stack}.  No
explicit data quality control has been performed (although only data
accepted for use in the survey has been used) and a month-to-month
re-binning of the data was require to place data on a common
wavelength scale.  Local error estimates are shown for random
realisations of the stacked data as before.  Stacks are created for
the data both with and without the PCA sky subtraction.  In order to
extend the experiment beyond the 219\,hours of data available using a
single fibre from each observational frame we also repeat the
experiment using 50 fibres from each frame, which pushes the effective
exposure time to beyond 10,000\,hours.

It is clear that the {\it dedicated sky fibre only} sky subtraction
again reaches a noise floor beyond which sky subtraction is not
improved.  This occurs at a greater time than for the single frame
experiment in Fig.~\ref{noPCA stack} likely due to the presence of
stronger correlated errors when using data from a single science
frame.  The stacked data which has been processed with the PCA
procedure continues to show the steady decline in local residual error
with no compelling evidence for a systematic noise floor.  The rate of
decline in the local error is however not N$^{-0.5}$ as would be
expected from purely Poisson noise considerations, but rather
N$^{-0.32}$, closer to the cubed root of exposure time.  There is of
course a small contribution to the Poisson noise from the intrinsic
source flux in each spectrum which, while small and randomised over
redshift in this experiment, is none zero.  No comparable duration
sky-only data set is available with which to avoid this contribution
at this time.

\subsection{Comparison with N+S survey observations}
The long duration experimental results from Fig.~\ref{PCA stack}
indicate that the PCA sky subtraction process applied to dedicated sky
fibres observational data produces excellent results.  We now ask how
these results compare to those obtained with the N+S process.  As
stated previously, the N+S process requires longer individual
integrations to reach a common signal-to-noise level in the data.  If
single fibres are allocated to each science target, the N+S observing
procedure requires $\times$4 the integration to reach the same
signal-to-noise level as a dedicated sky fibre observation (due to the
50\% duty cycle on source while still retaining the full 100\% of the
noise budget from the sky spectrum in both the A and B components of
the nod observation).  If a pair of fibres is allocated to each target
and cross-beam-switching (CBS) is used, then this becomes only
$\times$2 in exposure time.  We shall for the moment ignore the reduce
targeting efficiency implied by allocating two spectroscopic fibres to
each target.

Since N+S observations require this extend observation time with
respect to the dedicated sky fibre observations, we should therefore
compare the local residual error estimate of the N+S observation with
a dedicated sky fibre and PCA sky subtraction observation of the
equivalent overall exposure time, in order to determine the optimal
observation strategy.  This comparison is made in Fig.~\ref{PCA
stack2}.

For short duration exposures, the local error estimate for dedicated
sky fibre data and for N+S data is comparable, and consistent with the
Poisson limit (as demonstrated in Figs.~\ref{GAMA obj}-\ref{wigglez
ns3}). For the PCA processed observation, we assume the $\tau^{-0.32}$
rate of reduction in local error estimated from Fig.~\ref{PCA stack}.
For the N+S data, we apply the $\times$4 ($\times$2 if using CBS fibre
pairs at the expense of raw target numbers per observation) increase
in exposure time to two rates of reduction in the residual error.
Firstly we use the theoretical value of N$^{-0.5}$, secondly we use
the slower rate of N$^{-0.38}$ observed in Fig.~\ref{noPCA stack}.  We
wish to find the exposure time at which the local error is lower with
increasing exposure time when using the N+S observation
strategy\footnote{Clearly if the true rate of error reduction for N+S
observations is identical to that of the PCA observations,
N$^{-0.32}$, this situation is never achieved.}.  These values can be
calculated from the assumed gradients as shown in Fig.~\ref{PCA
stack2} and are found to be 47\,hours (6502\,hours) without the CBS
fibre pairs assuming the -0.5 (and -0.38) power of decline and
7\,hours (81\,hours) if CBS fibre pairs are used.

\section{Visual comparison of spectral quality}
\begin{figure}
\includegraphics[width=70mm]{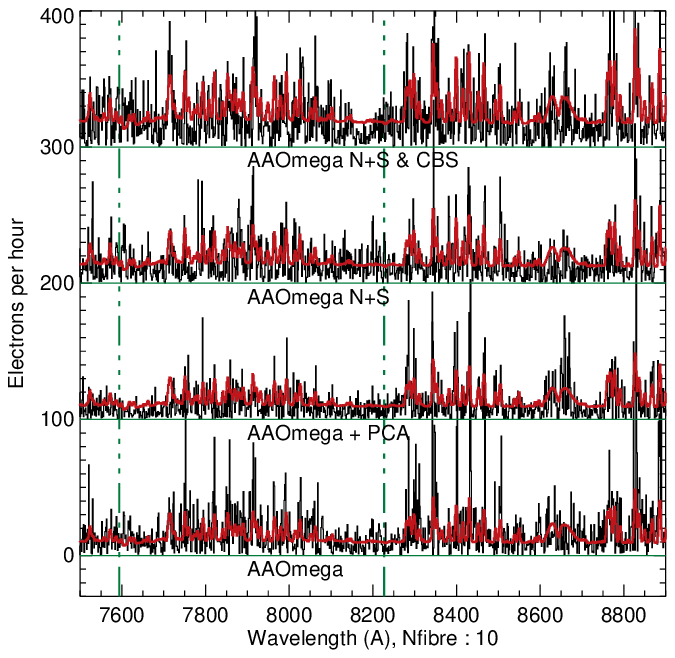}
\includegraphics[width=70mm]{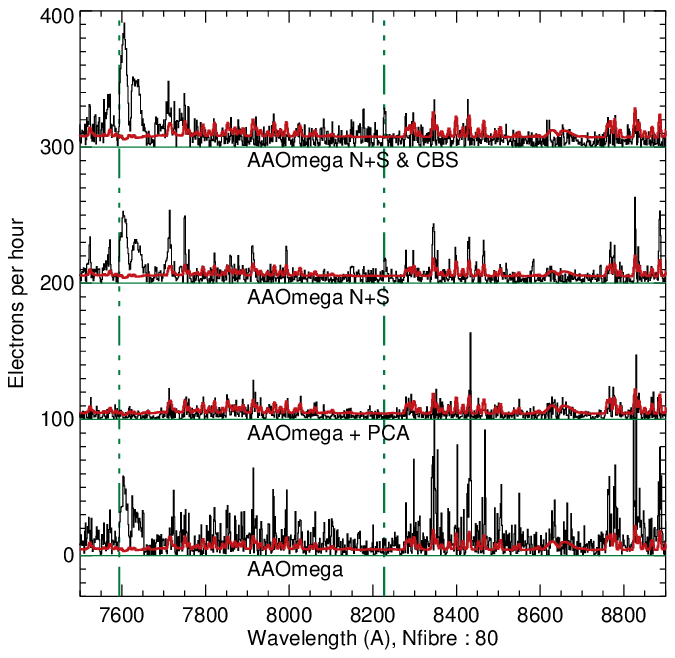}\\
\vspace{-0.5cm}
\caption{\label{stacked spectra ns} Stacked spectra are created using
the three sky subtaction methodologies under consideration.  The
moduls of the residual error in the stacked (continuum subtracted)
spectra is shown, along with the error estimate derived from variance
array propergation during data reduciton.  Vertical lines mark strong
Telluric bands which are not correctly addressd by the sky subtraction
process (the required multiplicative correction was not performed in
this case).}
\end{figure}

To illustrate the spectral quality of the various sky subtraction
methodologies we present the composite spectra, from the stacking
analysis using each of the processing methodologyies, in
Fig.~\ref{stacked spectra ns}.  The indicated number of spectra are
combined (using a clipped mean) after subtracting a smooth continuum
from each.  The modulus of the residual spectrum is shown in each
case, overlayed with the error array derived from the pixel variance
estimates during data reduction. In each case four data sets are
shown.  In the lower spectrum we see the 3x1200\,sec observation using
the dedicated sky fibres method. The second spectrum is the same frame
but with the PCA process applied to the data prior to stacking.  The
third and forth spectra are the N+S observations, the third does not
implement CBS fibre pairs while the upper spectrum does.  The N+S
frame was composed of 3$\times$2400\,sec observations and so the CBS
frame contains 2\,hours on source data, while the the other contains
only 1\,hour.

For the dedicated sky fibre observations the residual error is
significantly in excess of that expected from the data reduction
process and increasingly so as more spectra are combined (in agreement
with Fig.~\ref{PCA stack}).  The N+S data is broadly consistent with
the variance estimation but as is expected the level of this residual
error is higher due to the increased sky component of the N+S
observations.  The frame processed with the PCA sky subtraction is
visually by far the cleanest spectrum, supporting our previous
analysis.  It gives the greatest signal-to-noise data with the lowest
sky residual for the shortest on sky integration time.  Furthermore it
has considerably smaller observational overheads in comparison to the
N+S observations and maximises the fibre multiplex since there is no
need for cross-beam-switching.

An alternative assessment of the data quality is presented in
Fig.~\ref{stacked spectra}.  Here we create stacked spectra from the
219 WiggleZ observations from 2009.  In each case the spectrum is
presented with and without the PCA sky subtraction.  Insufficient N+S
observations are available to generate the comparison spectrum at this
time.  The suppression of systematic residuals in the spectra is
clearly visible in the PCA processed data.  No continuum subtraction
was applied to the fibres in Fig.~\ref{stacked spectra} and so the
weak source continuum from the high equivalent width emission line
targets of the WiggleZ survey is visible in the averaged spectrum.

\section{Quantitative survey analysis}
The most apparent benefit of the PCA analysis to the user, for
faint-object spectroscopy, is the reduction in systematic cosmetic
defects.  In order to provide a quantitative assessment of the
improvement a detailed simulation has been undertaken.  AAOmega survey
data was selected from six clear dark nights from the September 2008
observing campaign for the WiggleZ project.  This provided
85$\times$1200\,sec science frames.  As discussed previously, WiggleZ
targets are primarily high equivalent width faint ($r$(AB)$\sim$22)
emission-line sources hence each of the $\sim$354 science fibres
largely represents blank sky observation.  Additionally, each date
frame contains 25 fibre allocated to preselected blank sky positions
which are free from the low level Poisson noise contribution from the
faint continuum still present in fibres allocated to WiggleZ science
targets.  In what follows the blank sky fibre provide the idealised
test data set while the science fibres observations are used to
provide a larger sample size.

Each CCD frame was processed using the \texttt{2dfdr} software
environment and the standard data reduction recipes.  For each
individual reduced data frames (1200\,sec integration) a synthetic
spectrum was generated from a template early type galaxy.  The galaxy
template used was the intermediate age early type spectrum from the
GDDS survey \citep{gdds}.  The galaxies where distributed in redshift
between 0.7$<$z$<$1.3 which is the limit imposed by the rest frame
spectral data matched to the observed spectral range.  This redshift
range see the Calcium H+K absorption lines and the Balmer break
transit the strongest OH sky-line regions in the red AAOmega spectrum.
Spectra where generated for a series of source intensities scaled to
match count rates equivalent to apparent $I$(Vega)=20-24, with the
appropriate target Poisson noise included.

The PCA sky subtraction was then applied to each reduced science frame
and image stacks where created at each simulated magnitude level with
and without the PCA correction.  A redshift analysis was undertaken
with the fully automated mode of the \texttt{runz} cross-correlation
code typically used with data from the 2dF/AAOmega system.  The fully
automated analysis was chosen in preference to the more typical
semi-interactive methods more common to faint object programs
undertaken with AAOmega (e.g.\ \citet{Wake06,Ross06}) to avoid
subjective results.  For this idealised test case, only a single cross
correlation template was used, the same spectrum used in the
generation of the mock data.  The prior for the redshift distribution
was match to that of the mock test data input range.

\begin{figure}
\includegraphics[width=80mm]{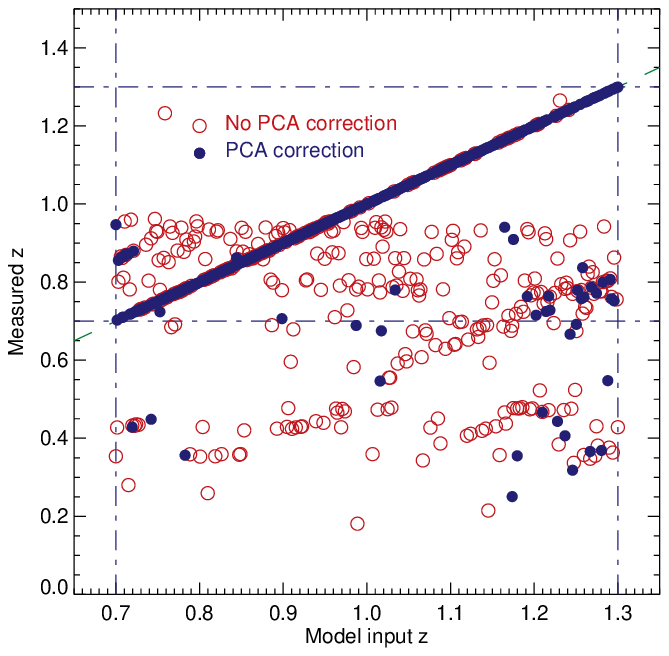}
\includegraphics[width=80mm]{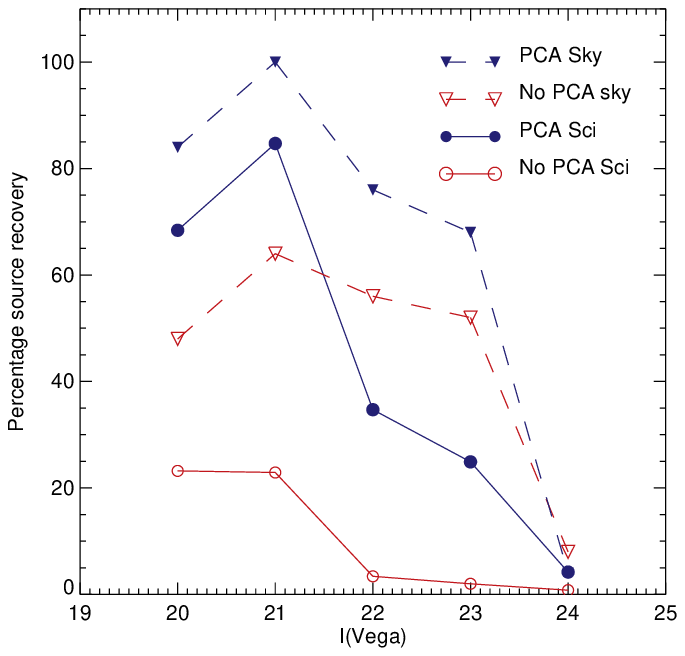}
\caption{\label{mockdata} The rate of successful redshift
determination is compared, with and without the PCA sky subtraction,
using a mock data set generated from 11.3\,hours of data
(85$\times$1200\,sec frames) from the WiggleZ survey.
\newline
Upper panel) The redshifts recovered by the \texttt{RUNZ} code in
fully automated mode are shown as a function of the input model
redshift.  Open circles show the results without the PCA correction,
filled symbols include the improved sky correction.  The data set
shown is for synthetic spectra with $I$(Vega)=21.
\newline
Lower panel) The rate of correct redshift identification is shown as a
function on the input model magnitude.  The simulations are performed
using the 354 science fibres, and also the 25 sky fibres available
across each frame.}
\end{figure}

The redshift analysis results are shown in Fig.~\ref{mockdata}.  The
upper panel indicates the measured redshift returned by \texttt{RUNZ}
as a function of input model redshift for the $I$(Vega)=21 simulated
data set for the full 11.3\,hour (85~frame).  With the PCA processed
data we recover the correct redshift for 85\% of simulated sources.
The misidentified spectra are preferentially at the highest redshifts,
where the $I$-band normalisation results in the lowest signal-to-noise
ratio spectra.  In the absence of the PCA sky correction, the success
rate is reduced to 25\% with a number of distinct failure modes
clearly present.

A comparison of the model spectra derived from blank sky fibre spectra
rather than those containing WiggleZ science targets demonstrates that
the residual noise contribution from the faint WiggleZ targets present
in the science spectra used for this analysis does contribute a
reduced efficiency for the test and will therefore contribute to the
slope measured for the sky error reduction rate seen in Fig.~\ref{PCA
stack}.

The lower panel of Fig.~\ref{mockdata} shows the declining recovery
rate for fainter model spectra.  At fainter input magnitudes, the
fully automated \texttt{RUNZ} code, which was not specifically tuned
for such a regime, does remarkably well even at $I$=23.  With the
planed upgrade of the red arm of AAOmega to a new generation {\it
high-resistivity} CCD (boosting quantum efficiency in the range
800nm-1$\mu$m by a factor of four) it will be possible for AAOmega to
continue to conduct efficent wide field surveys for early type
galaxies \citep{2SLAQ} out to at redshift z$\sim$1.

\section{Conclusion}
\label{conclusion}
We have demonstrated that the use of a Principle Components Analysis
sky subtraction with dedicated sky fibres observations in a manner
similar to that prescribed by \citet{Kurtz00} and \citet{Wild05} (and
subsequently implemented for the GAMA survey by \citet{Parkinson}),
provides excellent quality astronomical spectra over long duration
exposures.  For moderate length exposures (10-100\,hours on source)
the spectral quality is comparable to that obtained using the N+S
observational technique, and likely for significantly longer
exposures.  We therefore conclude that for fibre based multi-object
spectroscopy, in instances were high multiplex and observing
efficiency are critical, the PCA approach is superior to
nod-and-shuffle based techniques for moderate signal-to-noise
extragalactic survey spectroscopy.

\section*{Acknowledgments}
We wish to thank Scott Croom, Will Saunders and Karl Glazebrook for
invaluable discussion regarding the concepts discussed throughout this
work.  This investigation would not have been possible without the
hard work and dedication of the staff at the Anglo-Australia
Observatory, who designed, constructed, operate and maintain the AAT
and 2dF/AAOmega system with which the observations presented in this
paper where undertaken.

\bsp

\begin{figure*}
\includegraphics[width=85mm]{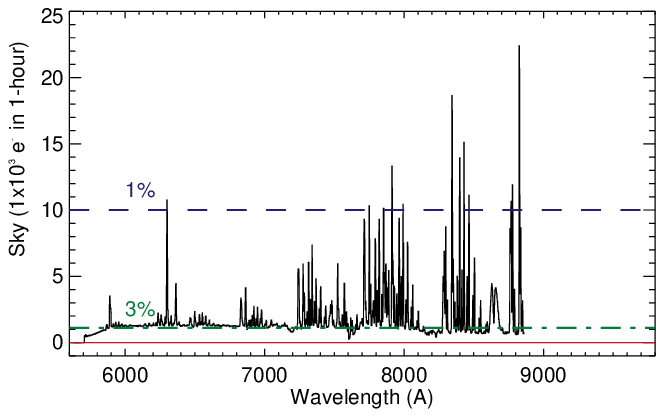}
\includegraphics[width=85mm]{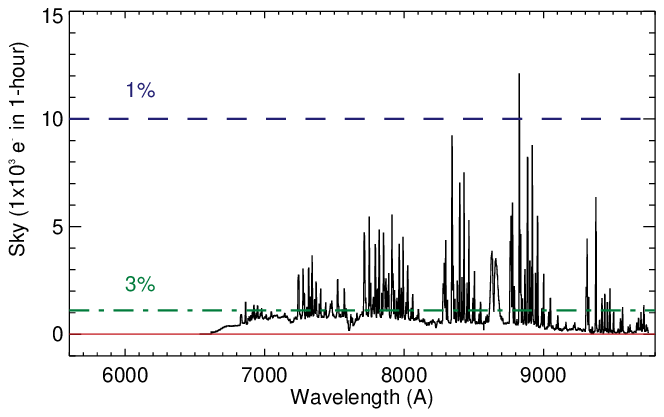}\\
\includegraphics[width=85mm]{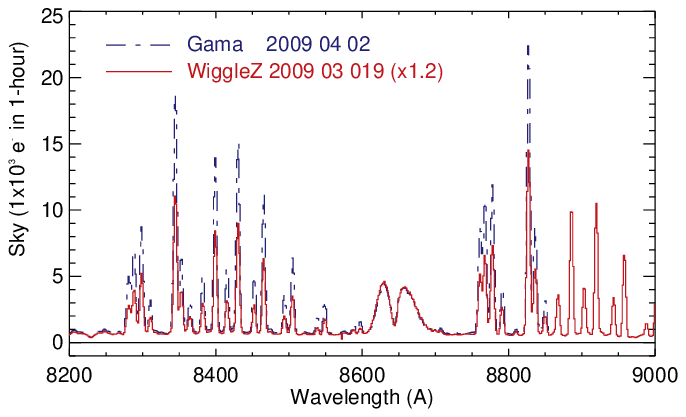}
\includegraphics[width=85mm]{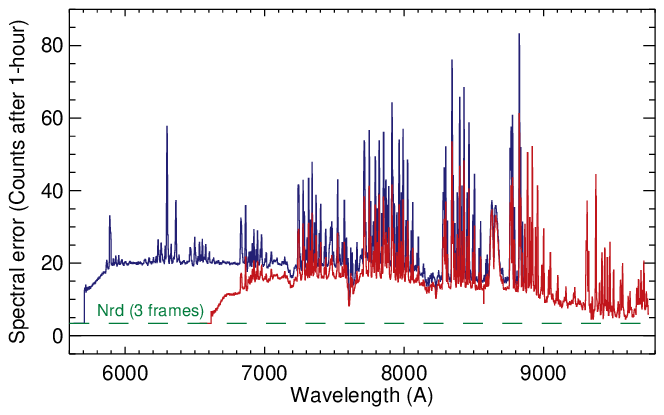}
  \caption{\label{skyspec} Representative sky spectrum are shown taken
  from the red arm of the AAOmega spectrograph.
\newline
  Upper panels -- The 5700\AA\ and 6700\AA\ dichroic beam-splitters
  have been used with the 385R VPH diffraction grating to cover the
  wavelength range 6000\AA\ $< \lambda <$ 9800\AA. The strong OH
  air-glow lines and the unresolved emission band of O$_{\rm2}$ are
  evident.  The spectra are gain and exposures time corrected to units
  of {\it electrons per hour}.  They are not flux calibrated and so
  line relative intensities are subject to the instrument response
  functions.  Horizontal lines marks the 1\% \& 3\% sky subtraction
  accuracy limits from Poisson noise.
\newline
  Lower panels -- Left) The overlap region of the two sky spectra are
compared.  The data were taken two weeks apart and a significant
variation in the OH sky-line scaling is evident.  The WiggleZ survey
spectrum has been scaled by a factor of 1.2 to match the O$_{\rm2}$
band, but the relative intensity of the OH lines remains higher in the
April data.  Right) The estimated residual error levels (1-$\sigma$
assuming purely Poisson noise dominated by the sky emission level) for
the two sky spectra are also shown, along with the expected
contribution from CCD readout noise assuming three independent
1200\,sec exposures.}
\end{figure*}

\begin{figure}
\includegraphics[width=85mm]{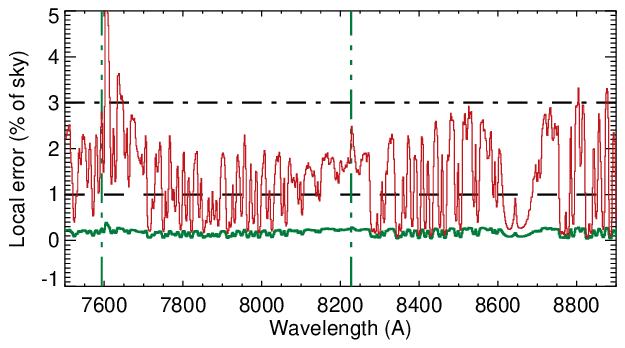}\\
\includegraphics[width=85mm]{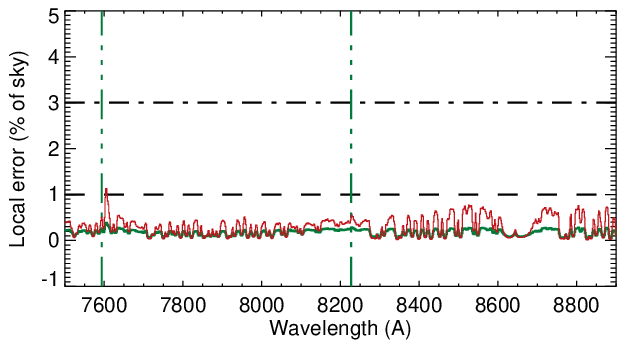}\\
\caption{\label{WiggleZ sky} Since the majority of targets from the
WiggleZ survey possess little continuum a stacking analysis can be
performed using all 352 science fibres from this frame.  This is shown
in the upper pane.  The presence of systematic defects is evident in
the rate of residual error reduction when comparied to the error level
expected from the variance arrays of the combined spectrum.
\newline
In the lower panel the stacking process is repeated after applying the
PCA correction to the data.  The local error is largely
indistinguisable from the Poisson error in this vissual
representation.  The associated residual error histograms
(\S~\ref{skysubdefinition}) are given in Fig.~\ref{WiggleZ hist}.}
\end{figure}

\begin{figure}
\includegraphics[width=70mm]{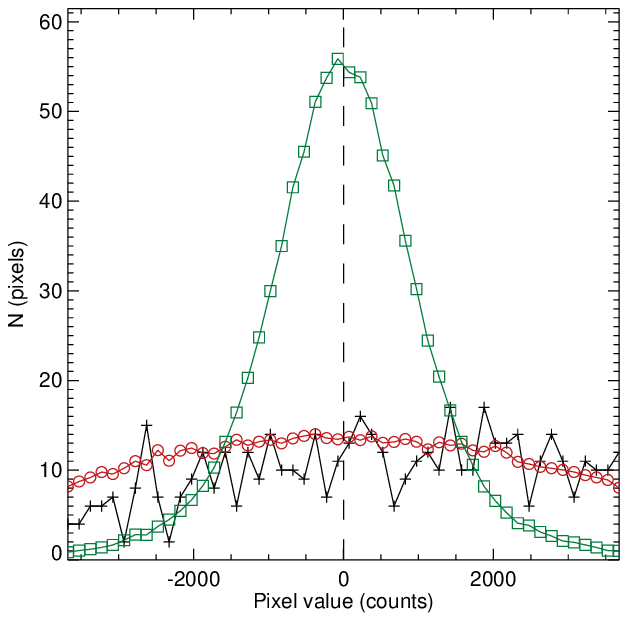}
\includegraphics[width=70mm]{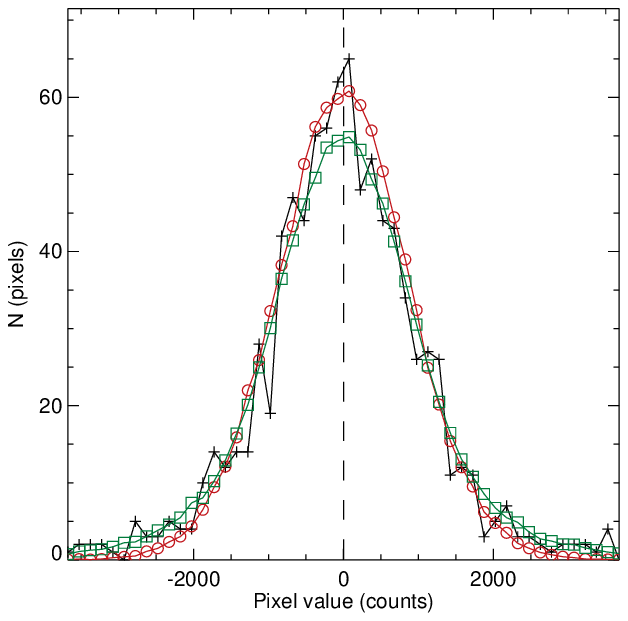}
\caption{\label{WiggleZ hist} The residual error histrograms for the
stacking analysis of for all 352 WiggleZ spectra from
Fig.~\ref{WiggleZ sky}.  The first is that for the 352 spectrum stack
without the PCA analysis.  The second is for the same data set but
with the PCA sky subtraction applied to the data prior to stacking.
The PCA process suppresses of the systematic error component and the
local error estimate is much closer to that expected from the spectral
variance estiamtes.}
\end{figure}

\begin{figure*}
\includegraphics[width=60mm]{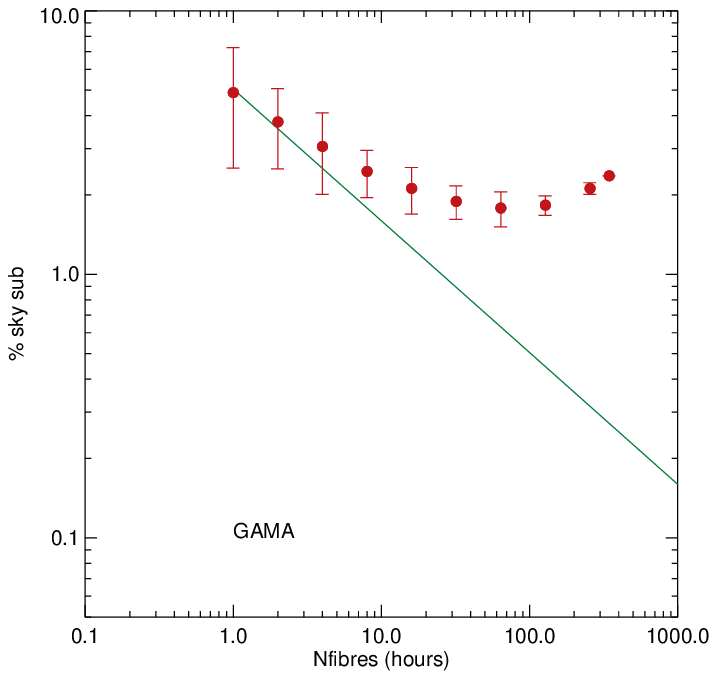}
\includegraphics[width=60mm]{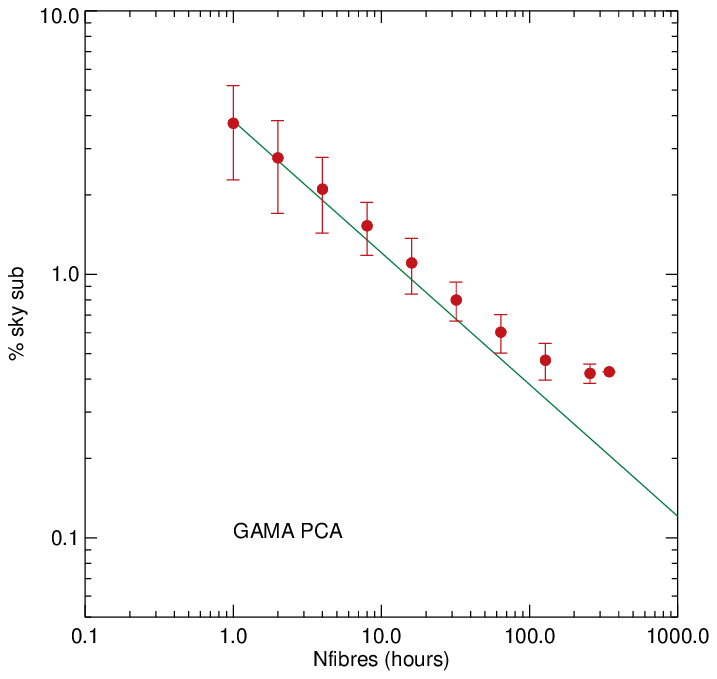}\\
\includegraphics[width=60mm]{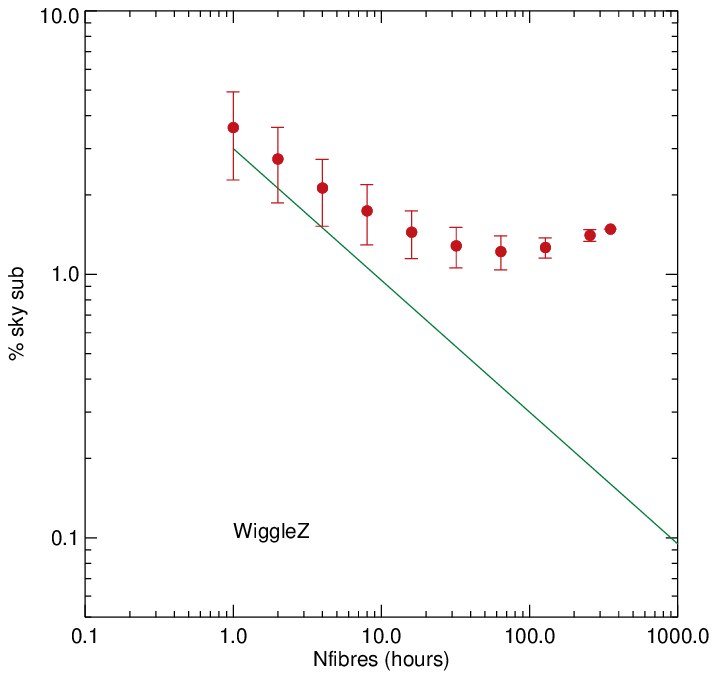}
\includegraphics[width=60mm]{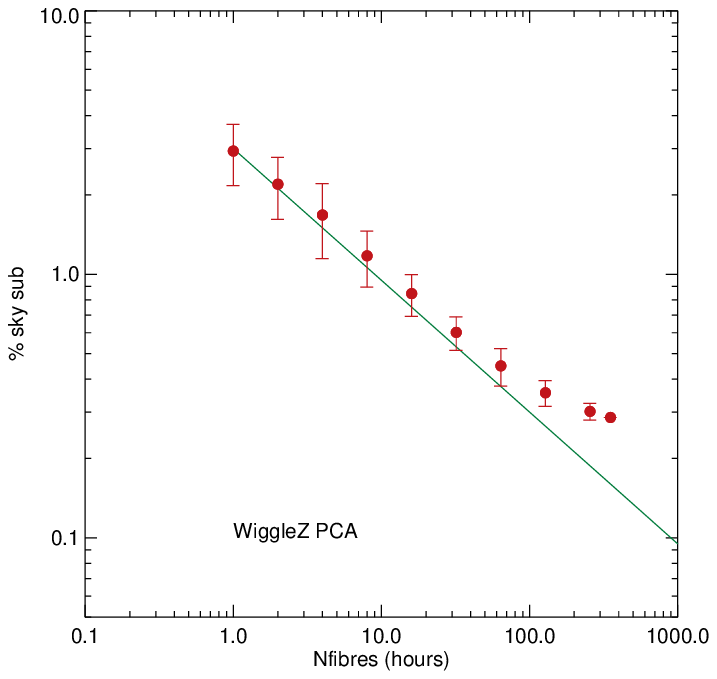}\\
\includegraphics[width=60mm]{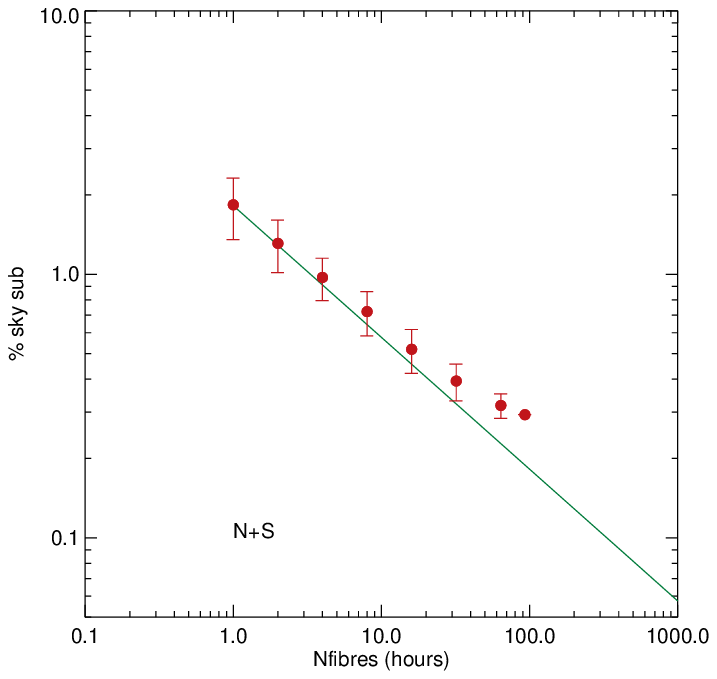}
\includegraphics[width=60mm]{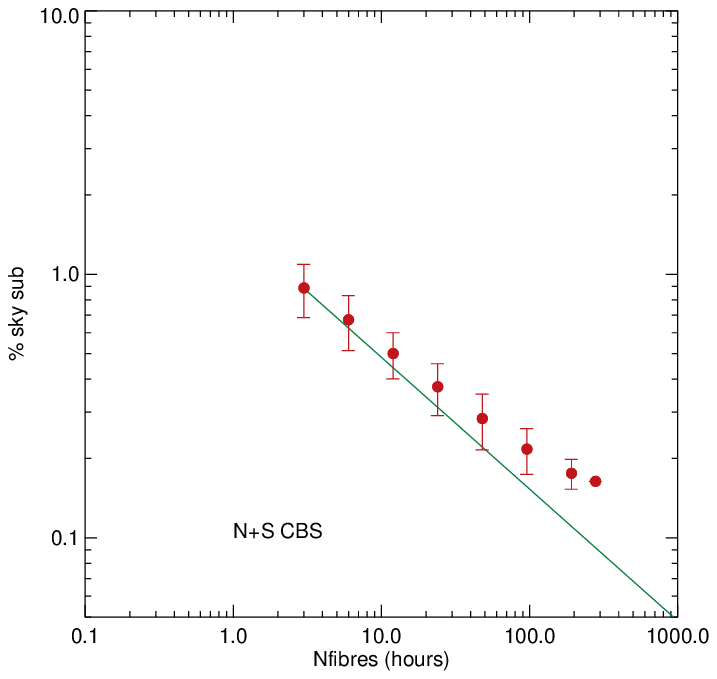}
\caption{\label{noPCA stack} In the absence of a large number of
independant data sets, long integration times can be simulated by
stacking the spectra from multiple fibres within a data set.  Local
residual error is plotted (as a percentage of the local sky intensity)
as a function of the number of fibres stacked in the three dats sets
considered.  The mean value $\pm$3-$\sigma$ rms scatter is shown for
100 random realisations of the stacked spectrum with each value of
Nfibres.  Only $\sim$400 fibres are avalibale in the GAMA and WiggleZ
frames.  This becomes $\sim$200 for the N+S observations.  The solid
line traces the $\sqrt{\rm{N}}$ decay expected from pure Poisson noise
in the data.}
\end{figure*}

\begin{figure}
\includegraphics[width=80mm]{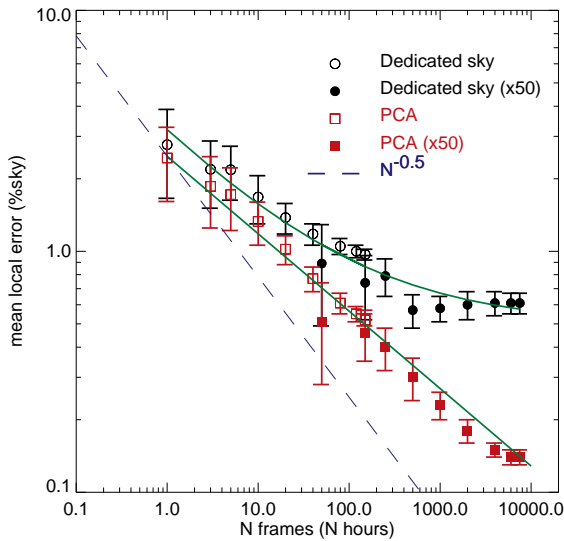}
\caption{\label{PCA stack} Some 219 independant frames are available
from the 2009 WiggleZ observing campaign with which to undertake a
stacking analysis.  The hollow points are constructed by combining a
single fibre from each of a N frames ($\pm$3-$\sigma$ rms scatter).
The filled points simulate a much longer exposure by including 50
fibres from each of N frames when making the stacked spectrum.  A
dashed line marks the $\sqrt{\rm{N}}$ rate of decline.}
\end{figure}

\begin{figure}
\includegraphics[width=80mm]{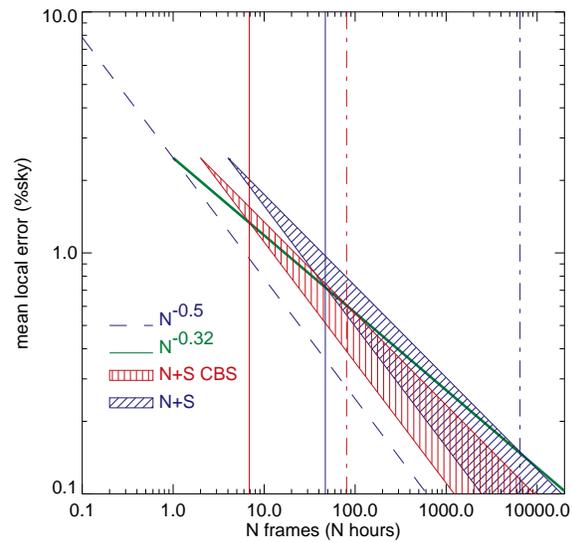}
\caption{\label{PCA stack2} The sky residual decline rate with
incearsed exposure time, for the PCA sky subtraction, is reproduced
from Fig.~\ref{PCA stack}.  Two additional zones are added to the
figure. Each zone shows the sky subtraction accuracy expected for N+S
observations assuming slopes of N$^{-0.5}$ (the theoretical limit) and
N$^{-0.38}$ (that observed in the observational test data presented in
figure Fig.~\ref{noPCA stack}.  The two zones have been shifted along
the horizontal axis (N frames, or rather exposures time) by the
multiplier required over standard dedicated sky fibres observations in
order to achive the same final signal-to-noise ratio in the spectrum
(with and witout CBS, i.e.\ x2 or x4).  It is not until the green
(PCA) line crosses these two boundaries that N+S observations become
{\it more efficent} than dedicated sky firbes observations.
\newline
The cross-over points indicated are 47\,hours (6502\,hours) for No CBS
with the slope of N$^{-0.5}$ (N$^{-0.38}$).  This becomes 7\,hours
(81\,hours) if CBS is implemented.}
\end{figure}

\begin{figure}
\includegraphics[width=80mm]{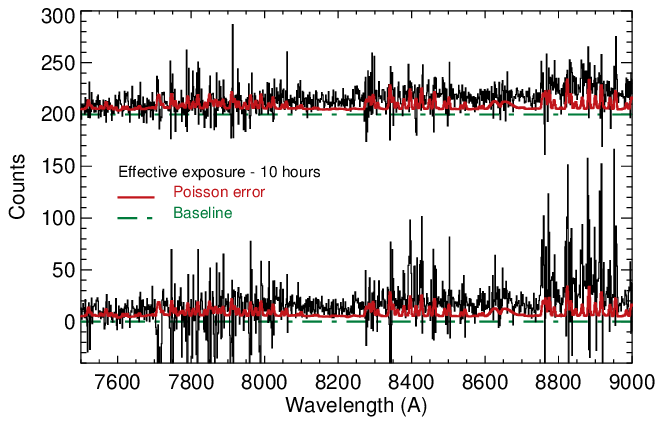}\\
\includegraphics[width=80mm]{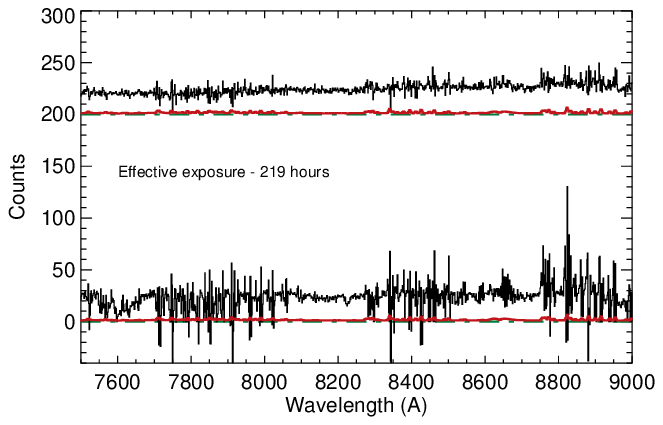}
\caption{\label{stacked spectra} Long duration exposures are generated
across the 219 frames available from the 2009 WiggleZ observing
campaign.  The indicated number of spectra are combined across a
number of independent frame.  Since the fibres used were placed on
different sources during each observation, there is no coherent
object signal present.  In each case the lower spectrum shows the
resulting spectrum without the PCA sky subtraction.  The correction
has been applied to the components of the upper spectra before
stacking, eliminating the systematic error residuals. The error array
from the data reduction process is also shown.  No continuum
subtraction was applied to the spectra prior to stacking and hence the
low level composite object continuum is visible.  The PCA processed
spectrum shows significantly smaller systematic residual errors,
consistent with Figs.~\ref{noPCA stack} \& \ref{PCA stack}.}
\end{figure}

\begin{table*}
\caption{\label{data sets} Details of the example AAOmega data sets
used in this analysis.}
\begin{tabular}{llll}
{\bf Sky subtraction mode} & {\bf Program} & {\bf Exp. details} & {\bf Observation Date}\\
\hline
\hline
Dedicated sky fibres & GAMA survey \citep{GAMA}     & 3 $\times$ 1200\,sec & 27 March 2008\\
Dedicated sky fibres & WiggleZ survey \citep{WiggleZ} & 3 $\times$ 1200\,sec & 25 March 2008\\
\hline
Nod-and-Shuffle & WiggleZ survey \citep{WiggleZ} & 3 $\times$ 2400\,sec $\times$ 3\,nights & 21, 25 \& 26 September 2008\\
\hline
Extended PCA test    & WiggleZ survey \citep{WiggleZ} & 3 $\times$ 1200\,sec & 2009\\
\hline
\end{tabular}
\end{table*}

\begin{table*}
\begin{center}
\caption{\label{skysubresults}Summary of sky subtraction accuracy
results.  For each data set, the predicted 5-$\sigma$ (per spectral
resolution element) $R$-band magnitude limits is indiacted}
\begin{tabular}{lllll}
{\bf Data set } & {\bf R}   & {\bf Poisson error}  & {\bf Local error}  &{\bf Exposure time details}\\
                & {\bf mag(Vega)} & {\bf \% sky}   & {\bf \% sky} \\
\hline
\hline
GAMA            & 22       & 3.53$\pm$1.17  & 4.67$\pm$0.60  & 1h on source (1h effective sky)\\
GAMA-PCA        &          & 3.53$\pm$1.17  & 3.76$\pm$0.46\\
\hline
WiggleZ         & 22       & 3.28$\pm$1.34  & 3.57$\pm$0.38  & 1h on source (1h effective sky)\\
WiggleZ-PCA     &          & 3.28$\pm$1.34  & 3.01$\pm$0.29\\
\hline
WiggleZ Deep       & 21.25    & 2.32$\pm$0.95  & 1.80$\pm$0.13 & 2h sky, 1h on source (2h effective sky)\\
WiggleZ Deep CBS   & 21.65    & 1.64$\pm$0.67  & 1.27$\pm$0.10 & 2h sky, 2h on source (4h effective sky)\\
WiggleZ Deep Full  &  22.25    & 0.95$\pm$0.40  & 0.87$\pm$0.07 & 6h sky, 6h on source (12h effective sky)\\
\hline
\end{tabular}
\newline
Note: for the N+S data, the sky is effectivly twice as strong, as
there is a noise contribution from both the on and off observing
positions.  For the CBS frames the sky is still double the open
shutter time, however the on source exposure time is also the full
open shutter time.
\end{center}
\end{table*}

\appendix
\label{Appendix}

\section{Sky subtraction methodologies}
\label{Appendix-methods}
The limiting sensitivity of a spectral observing system is ultimately
a question of the signal-to-noise ratio achievable for an given
observation.  In the faint source limit the Poisson-noise component
pertaining to the source photon arrival rate can be largely
ignored. Additionally one may assume an observational strategy has
been adopted such that noise injected by the detector system
(read-noise) may be discounted.  In what follows we shall also assume
no significant contribution from a thermal or scattered light
background.  Under these circumstances, at wavelengths in the range
700\,nm-1.7\,$\mu$m the limiting factor for system sensitivity becomes
the Poisson-noise associated with the foreground screen of emission
from the earths atmosphere.  On a moonless night, the dominant
component of this emission is the molecular emission bands from upper
atmosphere OH radicals, with an additional contributions from bands
due to O$_2$ (Ellis \& Bland-Hawthorn 2008). An example of this
OH-airglow spectrum, as seen by the red arm of the AAOmega
spectrograph, is shown in Fig.~\ref{skyspec}.  We will return to the
troublesome effects of scattered moon light later.

While this sky signature may be subtracted from science spectra, the
random noise component remains and can be suppressed only through the
collection of a greater quantity of photons, be that via a larger
telescope collecting area or an extended exposure time\footnote{An
important caveat being of course that increasing the source:sky photon
ratio via the use of a smaller aperture coupled to improved image
quality has an important effect.}.  In the sky-noise limit increased
depth is gained only at the square root of the increase in either.  In
the presence of even subtle systematic defects in the recorded
spectrum of a faint target, increasing the exposure time without end
will ultimately reach a noise floor limited by the presence of
systematics (Figs.~\ref{noPCA stack} \& \ref{PCA stack}).  The depth
attainable with any observing system is therefore limited by ones
control of sources of systematic error.

Since fibres were first used for astronomical spectroscopy (see
\citet{Hill88} for an historical perspective) a number of strategies
have been implemented to attempt to remove the sky signature which is
signed, in what often appears to be indelible ink, across every
spectrum.  A number of standard strategies have evolved over time; we
discuss each in turn below.

\subsection{Dedicated sky fibres}
The most common method for sky subtraction with fibre multi-object
spectrographs is to utilise \emph{dedicated sky fibres}.  In this
model, a number of fibres are allocated to previously determined blank
sky positions across the field of view.  This allows the observation
of the sky spectrum simultaneously with the science observation
removing the effects of temporal variations in the night-sky spectrum.

These sky spectra are processed alongside the science data and
combined to provide a {\it photon-noise} suppressed sky spectrum
(benefiting from the $\sqrt{\textsc{n}}$ noise suppression from
multiple independent samplings of the sky).  This combination creates
a \emph{master sky spectrum} effectively \emph{noise free} by virtue
of a high signal-to-noise with respect to the science observations.
To achieve this {\it noise free} sky spectrum, one must typically
combine 15-30 individual spectra.  The exact nature of the sum
depending on the observational requirements, and factors such as the
cosmic ray rate at the detector and the care with which the user has
chosen {\it blank} sky regions (a non trivial problem for observation
in some fields e.g.\ the galactic bulge).

While the use of the dedicated sky fibres prevents the act of
subtracting the master sky spectrum increasing the random noise (since
the master sky spectrum is essentially noise free), the subtraction
accuracy is at the mercy of a series of systematic defects as outlined
below.  These systematic effects limit the ultimate depth of
observations with the dedicated sky subtraction method, shifting the
signal-to-noise firmly off the $\sqrt{\tau}$ curve-of-growth one would
expect from a purely Poisson process of photon arrival from the target
and night sky.  The number of fibres one should dedicate to sky is a
topic of debate for many a survey program.  Clearly each fibre used
for blank sky is unavailable for observations of a science target
object and so directly reduces the multiplex advantage.  For AAOmega
observations at the AAT 20-25 fibres are typically allocated to blank
sky in a given configuration, $\sim$5\% of the full complement of
available science fibres.  It should be noted that this order of
overhead will likely remain for systems with many more fibres than
AAOmega.  The available fibres of such a system will likely be
separated across multiple spectrographs and for reasons of PSF
stability (outlined below) 15-25 fibres will likely be needed per
spectrograph.

The master sky spectrum is subtracted from all object spectra after an
appropriate scaling factor is derived to account for fibre relative
response variations.  Standard \emph{dome flat} frames typically
cannot be used to determine this relative response since it is not
usually feasible to provide a uniform intensity illumination to all
fibres across the field-of-view of the observing system.  For example
the quartz-halogen flat field lamp frames used with AAOmega
observations provide an excellent wavelength dependant relative
response fibre-to-fibre, but provide no total intensity normalisation
due to a strong illumination gradient in the flat field. This
illumination gradient results from the short path length ($\sim$2m)
between the light source and the diffuser screen.  For low resolution
sky limited AAOmega observations this scaling is usually derived from
measurements of the very night sky air-glow lines one aims to subtract
from within the sky and science target spectra.  The master sky
spectrum is then scaled to match the measured line intensities for
subtraction.

\subsubsection{Problems associated with dedicated sky fibre observations}
\label{Appendix-problems}
There are a number of issues which limit the sky subtraction accuracy
typically achieved with dedicated sky fibres.  The first five problems
considered below amount to systematic modification of the individual
input spectra via cross-contamination or the addition of a
non-astronomical component which largely invalidates the dedicated sky
fibres approach.  The remainder are concerned with the introduction of
small scale variations in the apparent profile of fibre images on the
detector, the Point Spread Function (PSF).

\subsubsection*{Continuum variations}
While there is little evidence on small angular scales for structure
in the night sky OH emission, at least at wavelengths shorter than
$<1$\,$\micron$, gradients in the continuum emission particularly
during {\it bright-of-moon} do significantly modify the observed sky
spectrum as a function of field position (certainly for instruments
with a wide field-of-view such as the $\pi$\,deg$^2$ of the AAOmega
system).  Indeed for very wide fields of view, spatial variations in
OH emission intensities may not in fact average out over the duration
of an exposure.\footnote{Scattered moon light from a poorly baffled
telescope superstructure would of course be particularly damaging. It
would be almost impossible to correct for such an additive systematic
error, since it will be highly variable both spatially and
temporally.} The first order effect of strong moon-light gradients in
the field will be that line and continuum spectral components scale
differently across the field and cannot be adequately accounted for
using a single scaling derived from a single component master
spectrum.

\subsubsection*{CCD fibre packing}
Perhaps the most mundane effect one must account for is simply the
fibre packing onto the CCD.  With CCD real-estate at a premium, there
is an ever present temptation to push optical fibres closer and closer
together in order to increase the multiplex for a given detector area.
For the AAOmega system at the AAT, the fibre pitch (separation between
adjacent fibres) is $\sim$10\,pixels, with the fibre profiles
projecting an approximately Gaussian FWHM$\sim$3.4\,pixels.  A tighter
pitch results in significant overlap (cross talk) between adjacent
fibres.  Close packing hampers scattered light (background) estimation
and is particularly problematic in the high contrast calibration
frames such as flat fields, where one requires high intensities to
reduce Poisson noise sensitivity, but which may compromise accuracy by
inducing significant cross-talk errors between fibres.  Improved
extraction methodologies, which better account for fibre-to-fibre
cross-contamination, are invaluable \citep{Sharp10,Bolton10}.

\subsubsection*{Flat field irregularities and scattered light}
Imperfect flat fielding within the spectrograph will introduce errors.
A common problem is imperfect removal of scattered light on the
detector system either from a poor accounting for the wings of
individual fibre profiles, or from low level diffuse
scattering and ghost events from the myriad optical elements within
the spectrograph light path (strong ghosts being of such grave
consequence that they have been removed by adequate baffling during
the detailed spectrograph design process of course).  Accurate removal
of scattered light from science and calibration frames, particularly
high signal level flat field frames, is essential for accurate sky
subtraction \citep{Sharp10}.

\subsubsection*{Pixel-to-pixel sensitivity variations}
Pixel-to-pixel sensitivity variations must be considered carefully.
Any high frequency variations in the final image will distort the PSF
of individual OH lines within the sky spectrum and modulate the sky
subtraction accuracy.  Ideally pixel-to-pixel sensitivity variations
should be multiplicative corrected via a {\it long-slit} flat field
process whereby a uniform (both in wavelength and spatial intensity
variations) source illuminates the detector and local intensity
variations are identified through dividing such a suitably exposed
frame by itself after moderate smoothing with a low pass spatial
filter.  For many observing systems the small amplitude of variations
found in the current generation of (low fringing) CCDs appear
adequately corrected by the fibre-flat-field process without resorting
to external \emph{long-slit} flat fields. Fringing in the CCD, due to
variations in the thickness of the device, is a related issue and for
a spectroscopic system can essentialy be treated as part of any
pixel-to-pixel sensitivity correction.  For the AAOmega system the
effect of using a long-slit flat field to correct for pixel-to-pixel
sensitivity variations is found to be measurable but not critically
significant.

\subsubsection*{Fibre fringing}
Any small air-gap between optical surfaces in the fibre feed will
introduce an etalon effect and produce interference fringing in the
observed spectrum of a fibre.  If this air-gap remains stable over
time, then the fringing amplitude modulation will be corrected via the
flat-fielding process and amounts only to an, undesirable, loss of
observing efficiently.  However, if the air-gap is unstable with time
the frequency of the interference fringing will change and
flat-fielding will no longer correct the signal modulation.  In this
case sky subtraction through the use of a master sky spectrum is
doomed to failure since the relative intensities of sky lines are
modulated by the fringing.  Fibres that exhibit fringing are therefore
of limited value for spectroscopy.  Careful preparation of all
air/glass surfaces, and where possible immersion of fibre faces, is
essential to guard against these catastrophic spectral defects.  Fibre
fringing has been a perennial problem in a small fraction of AAOmega
fibres.  The root cause has been traced to differential thermal
expansion properties of materials within each 2dF fibre button and a
mechanical solution, removing the differential expansion, will shortly
be implemented.

\subsubsection*{Optical system PSF variations}
Point Spread Function (PSF) variations within the spectrograph optics
are a serious limitation to the accuracy of sky subtraction.  The PSF,
a measure of the projected image quality at each point in the
spectroscopic camera, will typically have a radial dependence due to
the camera optics, as well as a wavelength variation.  This results in
a fibre-to-fibre wavelength dependant variation in the resolution
element across the detector.  Coupling this with manufacturing
tolerances on fibre diameter (typically a few percent variation
between fibres from different batchs, with somewhat smaller variations
from fibres constructed from a single draw) one realises that no two
fibres will project spectra with identical resolution characteristics,
and hence OH line profiles.

\subsubsection*{Wavelength calibration}
More mundane but related issues regarding the fidelity of the
wavelength calibration introduce further complexity.  Typically one
might expect to calibrate a well exposed arc lamp spectrum to an
accuracy of the order 0.1\,pixel, but variations in external factors
such as the $f$-ratio at which arc light is fed into the fibres for
calibration can reduce the internal relative accuracy of this
calibration to a point that compromises sky subtraction, even if the
global external solution is within tolerance.

The limited spectral (and spatial during the extraction process)
sampling of profiles compounds these errors once interpolation of
spectra onto a common working coordinate scale becomes necessary to
combine individual sky spectra or to perform the relative baseline
shift required to subtract the master sky spectrum from the observed
data (methods which limit or avoid this re-sampling are possible
(e.g.\ the b-spline fitting methodology used extensivly in the SDSS
pipeline processing software; see \citet{Bolton07}).  Increasing the
PSF sampling presents the usual trade-offs of adversely effecting
signal-to-noise and spectral coverage.

\subsubsection*{Focal ratio degradation}
Practical construction considerations also have an impact on
performance.  In slit mask systems, the impact of slit roughness
(primarily burrs from the manufacturing process, and non uniform slit
width etc.) have long been known and are dealt with via the use of
\emph{slit function} calibrations, usually from twilight flat field
frames.  One candidate for the fibre equivalent of slit roughness is
Focal Ratio Degradation (FRD).  For the multi-mode optical fibres
commonly used for astronomy, FRD results in a broadening of the
angular distribution of light as it emerges from the fibre with
respect to the distribution with which it was fed in.  This effect
modifies the \emph{width} of the fibre profile, changing the
resolution element, and more insidiously modifies the projected
profile shape of the fibre.  Hence the myriad sky lines in a set of
sky spectra will have subtly different shapes at the spectrograph due
to FRD.  Careful controls during the preparation of the fibres as part
of the fibre feed construction can reduce the impact of FRD and
improve throughput by preventing FRD losses from over filled optical
elements in the system (resulting from poor accounting for changes in
the focal ratio due to FRD).

\subsubsection{p-Cygni profiles}
The practical upshot of most of these limitations is a tendency for
sky subtraction residuals to present the classic {\it p-Cygni} line
profile at each and every night sky emission feature.  A slight
mismatch between the master sky spectrum and the individual sky
spectra within each object spectrum leads to over and under
subtraction of the sky lines as a function of wavelength.

A number of strategies exist for minimising the p-Cygni (or reverse
p-Cygni) residual profiles.  For example one may benefit from
employing a secondary wavelength calibration derived from the air-glow
lines themselves.  Typically this will not appreciably modify the
wavelength solution of any given spectrum, but can correct for small
irregularities at the fraction of a pixel level.  For the AAOmega
system, an iterative {\it optimal sky subtraction} routine was adopted
within the {\tt 2dfdr} data reduction package\footnote{Credit is due
to S.Croom for this significant development within {\tt 2dfdr}.}.
Under this scheme, the master sky spectrum is iteratively subtracted
from each object spectrum and the residual assessed while fitting for
line scaling, small velocity offsets (at the sub-pixel level) between
the template and the data, and employing a smoothing filter to account
for PSF variations.

With more data at hand, \citet{Wild05} implemented a Principle
Components Analysis (PCA) approach to the problem of correction of
poor PSF residual sky subtraction.  This methodology, first proposed
by \citet{Kurtz00}, has been successfully implemented with AAOmega
data at the AAT by both the WiggleZ \citep{WiggleZ} and GAMA
\citep{GAMA} survey projects.  It has been adopted as an option within
the AAOmega data reduction environment for moderate signal-to-noise
extragalactic spectroscopy.\footnote{Credit for this work is due
independently to PhD students Emily Wisnioski (WiggleZ) and Hannah
Parkinson (GAMA).}

\subsection{Nod-and-shuffle}
The \emph{nod-and-shuffle} (N+S) observing technique was first
deployed at the AAT by \citet{Glazebrook01} for use with the LDSS
slit-mask system, building on the earlier work of \citet{Cuillandre94}
and the {\it Va et Vient} technique.  The premise of N+S observing is
to perform quasi-continuous observations of both target and blank sky
(adjacent to the target) through the identical light paths, hence
preserving the temporal and optical structure of the {\it on} and {\it
off} target sky spectra.  The principle is identical to that of beam
switching, but achieves a high chopping frequency (30\,seconds to
2\,minute dwell times) without incurring the onerous CCD read-out
times and read-noise associated with classical beam switching at this
frequency.

One therefore requires an observational scheme whereby charge can be
accumulated on the CCD for multiple independent observations, totaling
an observational time of the order 20-40\,minutes (sufficient to reach
the sky limited regime for observation with the low resolution AAOmega
gratings of R$\sim$1350), and then readout the entire observation in a
single (2\,minute) readout.  This mitigates the impact of readout
noise, while still allowing independent quasi-continuous observations
of target+sky position pairs.  Much has already been written about the
mechanics of CCD charge shuffling
\citep{Cuillandre94,Glazebrook01,gdds}, here we merely note that
charge accumulated in one pixel of a CCD can be somewhat arbitrarily
moved across the CCD, essentially instantaneously, with no loss of
signal integrity.

The CCD charge transfer phenomenon is fundamental to the basic
operation of a CCD, with all charge accumulated in an ordinary
observations essentially being charge \emph{transferred} across the
CCD to the readout register(s) before readout.  Prolonged experiments
indicate that, with modern CCDs, charge transfer between pixels is
essentially loss-less (a charge transfer efficiency of 100\%) at least
in the limit of low signal levels (high-signal high-frequency data
leaving some unfortunate charge transfer inefficiency residuals).

Furthermore, while a full CCD readout may take 2\,minutes,
transferring charge around the CCD is practically instantaneous.
Hence provided sufficient \emph{storage space} is available on the
CCD, one can follow the nod-and-shuffle procedure and accumulate
charge with the telescope at multiple pointings before performing a
single CCD readout.

\subsubsection{Limitations of Nod-and-Shuffle}
There are a number of immediate limitations which constrain the
utility of the {\it nod-and-shuffle} process.\\

\vspace{-0.25cm}\noindent{\it Duty cycle -} There is a 50\% on-source
efficiency hit since 50\% of the time each fibre is pointing at blank
sky.  This can be mitigated through cross beam switching (CBS) as
outlined below.\\

\vspace{-0.25cm}\noindent{\it Fibre masking -} For the 2dF/AAOmega
system, only 50\% of fibres are available (using the classic N+S
technique) since the remaining fibres must be masked off to provide
the storage area needed on the CCD for the charge shuffling.\\

\vspace{-0.25cm}\noindent{\it Increased Sky noise -} Since Sky is
observed in the {\it on} and {\it off} position and then a simple
image subtraction performed, there is no $\sqrt{\textsc{n}}$
suppression of noise in the {\it sky} spectrum as was the case for
{\it dedicated sky fibres} observations, hence there is a $\sqrt{2}$
increase in sky noise per spectrum.\\

\vspace{-0.25cm}\noindent{\it Read noise -} There is a $\sqrt{2}$
increase in effective CCD readout noise contribution due to the
subtraction of the B position spectrum, which can be significant for
shorter observations.\\

\vspace{-0.25cm}\noindent{\it Telescope overheads -} There will be a
telescope settling time overhead for each telescope nod.  Usually
telescope slew time is insignificant in comparison to the settling
time.  The time scale for variations in the night sky emission is
known to be short \citep{Glazebrook01}.  Experimentation with AAOmega
{\it nod-and-shuffle} indicates the best subtraction is achieved with
dwell times $<$30\,sec.  However, this incurs a significant overhead
in terms of telescope slew and settle time and an dwell time of
60\,sec is more typical.  Dwell periods in excess of 90-120\,sec are
found to adversely effect data quality.\\

\subsection{Mini-shuffling}
\label{mini-shuffle}
One clear limitation of the basic {\it nod-and-shuffle} strategy is
that one requires significant \emph{storage area} on the CCD in which
to hold charge-shuffled spectra during the alternate cycle of each
observation.  With the 2dF/AAOmega system this is provided by masking
alternate fibres from the spectrograph slit.  This is physically
achieved through the use of an opaque mask attached to the 2dF top
end.  This occults the light to alternate fibres, which have been left
in the \emph{parked} position around the edge of the 2dF field plate.
Attaching the mask is a trivial process, but it does require that the
telescope be parked at prime focus access between observations, which
adds 5-15\,minutes to the observing cycle (depending on the location
of targets on the sky with respect to the due north prime focus access
position).

With a high quality PSF within the spectrograph, one is at liberty to
reduce the fibre-to-fibre spatial separation (the fibre pitch) on the
CCD.  In the idealised situation, each fibre would be separated by
many multiples of the FWHM of the PSF, however tighter packing is
required to make reasonable use of the available CCD real-estate.  For
the AAOmega system, a fibre pitch of $\sim$10\,pixels is implemented,
for a fibre PSF of $\sim$3.2-3.5 pixels (AAOmega exhibits a broadened
PSF in the far red, a consequence of the greater penetration of
photons into the CCD before detection at these wavelengths, which
becomes significant due to the fast $f\sim1.3$ beam of the AAOmega red
camera).  Hence fibres are separated at the $\ge$1$\sigma$ level,
allowing a straight forward (intensity weighted) pixel summation
across each fibre profile during extraction, with minimal
fibre-to-fibre cross-talk \citep{Sharp10}.

With detailed modeling of the spectrograph PSF, one can implement an
\emph{optimal extraction} algorithm which simultaneously extracts the
profile information for overlapping fibre profiles, allowing fibres to
be packed more densely onto the CCD while limiting cross-talk
\citep{Sharp10,Bolton10}.  The optimal extraction routine implemented
for AAOmega within the \texttt{2dfdr} data analysis suite, while
significantly slower than a simple weighted summation has be shown to
produce $\sim$1-2\% cross talk residuals even for fibres with a pitch
$<1.5\times$FWHM \citep{Sharp10}.

Once tight packing of the spectra onto the CCD becomes practical, it
is possible to perform nod-and-shuffle observations with the full
fibre complement.  The AAOmega system was indeed designed with this
mode of operation in mind.  This {\it mini-shuffling} concept allows
all 400 AAOmega fibres to be used in a single charge shuffling
observation.  With a fibre FWHM profile 3.2 pixels, and a fibre pitch
of 10\,pixels, the interfibre gap is adequate to provide the storage
zone for B position pixels if an 800 fibre optimal extraction is then
implemented to account for the inevitable cross-talk resulting from a
fibre separation of $\le$1-$\sigma$ \citep{minishuffle}.

\section{Comparative efficiency of strategies}
\label{efficeny}
In making a decision on which observing strategy to pursue for any
given program, one must make a number of base-line assumptions with
regards to one's program requirements, and in addition consider a
number of mitigating factors.

\subsection{Baseline assumptions}
{\noindent\it FoV -} One requires the large field of view common to
fibre fed spectroscopic systems, and observations with a standard long
slit or slit-mask system cannot achieve the desired goal.\\

{\vspace{-0.25cm}\noindent\it Target density -} Sufficient targets are
available per field of view to utilise all available fibres (maximise
the multiplex), i.e.\ raw target numbers are a significant factor to
the experiment.\\

{\vspace{-0.25cm}\noindent\it Figure of merit -} Signal-to-noise and
target number are the figure of merit, not some more abstract (yet
important) measure of spectral fidelity.\\

{\vspace{-0.25cm}\noindent\it Fibre availability -} Assume no fibres
are required for calibrations.  i.e.\ we ignore the 5-7\% of fibres
usually allocated to simultaneous sky observations.  With the AAOmega
system this number is typically 25 fibres.  For large multiplex
systems we note that due to PSF variations etc., this percentage will
likely stay constant, requiring a similar number of fibres be
allocated to sky per unit spectrograph rather than per telescope
pointing.\\

{\vspace{-0.25cm}\noindent\it Sky limited observations -} The
observing strategy is assumed to move one out of the read-noise limit,
i.e.\ the error budget is dominated by {\it Poisson-noise} from the
sky background.  When limited by the arrival rate of photons from
one's target source, one is not likely to require the sky subtraction
accuracies provided by nod-and-shuffle observations in order to remove
one's background signal.

\subsection{Signal-to-noise implications}
The combined effect on the signal-to-noise ratio, given the baseline
assumptions, of observing with the nod-and-shuffle observing mode is a
2$\sqrt{2}$ reduction in the signal-to-noise ratio achieved for a given
integration time.  This reduction comes from the following sources.\\

{\noindent\it Exposure time -} A factor of 2 is lost in exposure time
since half of the time must be spent on sky. This is a $\sqrt{2}$ loss
in signal-to-noise.\\

{\vspace{-0.25cm}\noindent\it Target multiplex -} A factor of 2 is
lost in the number of targets that are observed since one can only use
half the fibres as the remaining half must be masked in order to
provide the storage area into which charge can be shuffled.  Therefore
to recover the same number of targets, one must observe two target
sets for half of the exposure time each, a $\sqrt(2)$ signal-to-noise
loss.\\

{\vspace{-0.25cm}\noindent\it Effective increased sky brightness -}
There is a factor of two increase in the effective sky background,
with respect to the object's signal, due to the presence of sky
emission in both the A and B positions.  The noise component from the
sky spectrum can not be suppressed as is the case with dedicated sky
fibres.  Essentially the sky is observed for the full exposure
duration while the target is only observed for half of this time.
This doubling of the sky signal is a further $\sqrt{2}$ effect in the
SN calculation.

\subsection{Mitigating factors}
There are a number of strategies which can be used to minimise the
impact of {\it nod-and-shuffle} on the loss of signal-to-noise ratio.
When all can be applied the loss in signal-to-noise can be reduced
from 2$\sqrt{2}$ to $\sqrt{2}$.\\

{\noindent\it Low source density -} In the limit of a reduced number
of valid targets per field, the loss from losing half of the fibres is
removed.\\

{\vspace{-0.25cm}\noindent\it Mini-shuffling -} The mini-shuffling
strategy outlined in section \ref{mini-shuffle} can be used to allow
the full fibre multiplex to be used.\\

{\vspace{-0.25cm}\noindent\it Cross-beam switching -} In the limit of
a reduced number of valid targets per field, two fibres can be
allocated per target to remove the 50\% on-source duty cycle effect.
Again, the mini-shuffling strategy allows the maximum multiplex.\\

The increased sky background cannot be mitigated, and so the
$\sqrt{2}$ signal-to-noise reduction per unit time from this component
remains.  However, this is a random error and not a systematic effect
and so scales with the square root of an increased exposure time.

In circumstances where a gain is found from the N+S technique the
observing strategy when employed with a fibre multi-object
spectrograph can deliver sky subtraction at the Poisson limit over a
wide field of view and with a high multiplex.  Our recommended
observing strategy for using N+S with the AAOmega facility at the AAT
is as follows:\\

\vspace{-0.25cm}\noindent{\it Configure} the target field using either
standard nod-and-shuffle with fibre masking (200 fibres) or using
mini-shuffling (400 fibres).\\

\vspace{-0.25cm}\noindent{\it Allocate} fibres to targets in pairs
using cross-beam-switching (CBS) to alleviate the 50\% on source duty
cycle.\\

\vspace{-0.25cm}\noindent{\it Observe} in blocks of 20 pairs of A/B
telescope nods, with 60\,seconds dwell per position, for individual
exposures of 40\,minutes per exposure.\\

\vspace{-0.25cm}\noindent{\it Perform} three/four such integrations
per 2dF fibres configuration in order to avoid losses from atmospheric
perturbations on target positions (due to differential atmospheric
refraction across the field).\\

\vspace{-0.25cm}\noindent{\it Telescope nods} should be performed
using 60-120\,arcsecond offsets to minimise telescope slew time while
allowing sufficient separation between A and B positions on the 2dF
field plate to allow CBS fibre pairs to be allocated.\\

\vspace{-0.25cm}\noindent{\it Gradients} in the pattern of variation
in the OH sky lines can be mitigated by observing in the classic ABBA
pattern.  This has the added advantage of removing the requirement for
a significant number of intermediate telescope nods.\\

\vspace{-0.25cm}\noindent{\it Telescope nods} should be performed in
the Declination (north-south) axis to minimise telescope settling time
for an equatorial telescope such as the AAT.

\label{lastpage}

\end{document}